\documentclass[12pt]{JHEP3}
\pdfoutput=1
\usepackage{amsmath,epsfig}
\usepackage{amssymb,amsfonts}
\usepackage{latexsym}
\usepackage[latin1]{inputenc}
\usepackage{tocvsec2}
\usepackage{subeqnarray}
\usepackage{xcolor}

\usepackage{graphicx}
\usepackage{longtable}

\relax
\renewcommand{\theequation}{\arabic{section}.\arabic{equation}}
\def\be{\begin{equation}}
\def\ee{\end{equation}}
\def\bea{\begin{eqnarray}}
\def\eea{\end{eqnarray}}
\newcommand\fverb{\setbox\pippobox=\hbox\bgroup\verb}
\newcommand\fverbdo{\egroup\medskip\noindent%
                        \fbox{\unhbox\pippobox}\ }
\newcommand\fverbit{\egroup\item[\fbox{\unhbox\pippobox}]}

\newcommand{\bear}{\begin{eqnarray}}

\newcommand{\eear}{\end{eqnarray}}

\newcommand{\bsea}{\begin{subeqnarray}}
\newcommand{\esea}{\end{subeqnarray}}
\newbox\pippobox

\def\6{\partial}

\def\a{\alpha}
\def\g{\gamma}

\def\pa{\partial}

\def\m{\mu}
\def\n{\nu}

\def\t{\theta}
\def\sp{\;\;\;,\;\;\;}

\def\sq
\def\a{\alpha}
\def\b{\beta}

\def\hri#1#2{\href{http://arxiv.org/abs/#1}{[ArXiv:#1]#2}}
\def\hre#1#2{\href{http://arxiv.org/abs/#1/#2}{[ArXiv:#1/#2]}}

\def\d{\delta}

\newcommand{\al}{\alpha}
\newcommand{\ga}{\gamma}

\newcommand{\da}{\delta}

\newcommand{\ud}{\mathrm{d}}

\newcommand{\half}{\frac{1}{2}}
\newcommand{\e}{\mathrm{e}}

\def\e{\epsilon}

\allowdisplaybreaks[3]

\title{Quantum critical lines in holographic phases with (un)broken symmetry}
\author{\large B. Gout\'eraux$^{a,b,c}$ and E. Kiritsis$^{b,c}$\\
~\\
~\\
$^a$\href{http://www.nordita.org}{Nordita}, KTH Royal Institute of Technology and Stockholm University\\
Roslagstullsbacken 23, SE-106 91 Stockholm, Sweden\\
~
\\
$^b$ \href{http://www.apc.univ-paris7.fr/APC_CS/}{APC}, AstroParticule et Cosmologie, Universit\'e Paris Diderot, CNRS/IN2P3, CEA/Irfu, Observatoire de Paris, Sorbonne Paris Cit\'e, 10 rue Alice Domon et L\'eonie Duquet, 75205 Paris Cedex 13, France;
~\\
~
\\
$^c$ \href{http://hep.physics.uoc.gr}{Crete Center for Theoretical Physics},
Department of Physics, University of Crete, 71003 Heraklion, Greece.
\\\\
E-mail: \email{blaise@kth.se}, \href{http://hep.physics.uoc.gr/~kiritsis/}{http://hep.physics.uoc.gr/~kiritsis/}
}

\preprint{CCTP-2012-24, NORDITA-2012-101}

\abstract{All possible scaling IR asymptotics in homogeneous, translation invariant holographic phases preserving or breaking a U(1) symmetry in the IR are classified.  Scale invariant geometries where the scalar extremizes its effective potential are distinguished from hyperscaling violating geometries where the scalar runs logarithmically. It is shown that the general critical saddle-point solutions are characterized by three critical exponents ($\theta, z,\zeta$). Both exact solutions as well as leading behaviors are exhibited. Using them, neutral or charged geometries realizing both fractionalized or cohesive phases are found. The generic global IR picture emerging is that of quantum critical lines, separated by quantum critical points which correspond to the scale invariant solutions with a constant scalar.}

\begin{document}

\section{Introduction}
\label{intro}

Effective holographic theories (EHTs) and their parametrization have been advocated
as a useful and powerful tool to analyze the phase structure of holographic strongly
coupled theories following the Wilsonian philosophy and classify all universality classes of holographic (quantum) critical behavior, \cite{cgkkm}.
In QFT, the space of all theories is organized by the Wilsonian paradigm in terms of fixed points, that correspond to scale invariant QFTs and flows among them.
The first step therefore is to classify scale invariant theories (SITs). Their properties determine in a concrete and unique way a local chart in their neighborhood.
Relevant and irrelevant operators indicate the directions in which one can flow out of or into such SITs. Such neighborhoods are known as scaling regions, and their properties can be analyzed in
perturbation theory provided that the SIT is solvable.

In QFT we know the basic Wilsonian rules, although the classification and solution of non-trivial SITs is an open problem in all dimensions including $d=2$,
 where the largest set of them are known.
The final part in organizing the space of QFTs is to describe the flows between SITs, and in particular to answer the question: where does a particular RG flow end up?
The generic answer to such a question is not known in all dimensions. Exceptions to this are the rare  cases where the full flow is perturbative, special examples in $d=2$ where the full flow is exactly solvable,  or cases where the endpoint, although at strong coupling, can be guessed from symmetry arguments and anomaly matching.

In this patch-wise construction of the space of QFTs, the geometry,  as was first introduced in \cite{za}, is locally well-defined.
However the structure of the full space is elusive as it contains singular and orbifold points, patches with different dimensions, and boundaries that are not accessible to analysis.
It is not known whether there are disconnected components in this space.
In a very concrete sense this looks very similar to what the space of vacua of string theory seemed to be like.
The AdS/CFT correspondence has made that similarity plausible at least for subspaces
of the two spaces.

The program of EHTs is addressing a similar coordinatization of the space of QFTs that have a semiclassical gravity dual and are at strong coupling. Following a similar strategy, we would like to know
what the scale invariant theories  are, how to characterize their scaling regions, and finally how to connect them with flows.
The answer to such questions, involving AdS vacua corresponding to Lorentz-invariant CFTs, is more or less similar to the Lorentz-invariant QFT answer.
However here there are several other cases of scaling behavior not linked to AdS that have not been well understood. The simplest example of this is Lifshitz scale symmetries, \cite{Li}
and Schr\"odinger symmetries, \cite{schr} while more generic cases are theories with generalized
 Lifshitz invariance and hyperscaling violation, \cite{cgkkm,gk,sachdev,dong} and the associated
Schr\"odinger cousins \cite{kim}. The analogous problem in QFT is the classification of non-Lorentz invariant scaling theories, \cite{lv}.
More complicated non-Abelian scaling symmetries (according to the Bianchi classification) have been recently addressed, extending the list of QC universality classes, \cite{kachru-bia,kasa}.

In \cite{cgkkm,gk} the most general quantum critical behavior of the metric with Abelian scaling symmetry has been determined and shown to be realizable in a class of EHTs involving also a U(1) gauge field (with unbroken gauge invariance) and a scalar. It is characterized by a Lifshitz exponent $z$ ($z=1$ corresponding to standard Lorentz invariance) and a hyperscaling violation exponent $\theta$ which controls the departure of the physics from naive scaling, \cite{cgkkm,gk,sachdev,dong,Edalati:2012tc}.
In this paper we will address a larger class of EHTs which realizes U(1) symmetry breaking with a complex scalar order parameter. We will give a classification of all possible scaling behaviors that can appear in that context. We will also argue that this class of EHTs captures the generic behavior
in most complicated cases. The surprising outcome will be that generically the IR asymptotics in symmetry breaking contexts are scale invariant, and therefore
there are gapless modes on top of those that are natural to expect in cases of spontaneous symmetry breaking.

The class of Effective Holographic Theories we will consider in this paper are described by the following action,
\be
 S_M=M^{2}\int d^{4}x\sqrt{-g}\left[R-{1\over 2}(\partial\phi)^2+V(\phi)-{Z(\phi)\over 4}F^2-{W(\phi)\over 2}A^2+X(\phi)F^*F\right].
\label{1}\ee
We will set the dimension of space-time to be four, and therefore the dual QFTs live in 2+1 dimensions.\footnote{However all the results generalize straightforwardly to higher dimensions.}
They involve three basic fields, the metric $g_{\m\n}$ dual to the stress tensor, a gauge field $A_{\m}$ dual to a U(1) current and a scalar $\phi$.
For the class of solutions that we will be considering in our paper, namely homogeneous, translation-invariant solutions without magnetic fields, the last term is not important.
 We will therefore drop it from now on.\footnote{Such a term is extremely important when physics in magnetic fields is discussed, and is also at the source of density wave instabilities, \cite{ooguri,donos}.} We give the field equations and the metric Ansatz in appendix \ref{app:Ansatz}.

When the mass term for the gauge field vanishes $W(\phi)=0$, U(1) gauge invariance is intact and the dual U(1) current is conserved. The action in that case
 \be
 S_0=M^{2}\int d^{4}x\sqrt{-g}\left[R-{1\over 2}(\partial\phi)^2+V(\phi)-{Z(\phi)\over 4}F^2\right]
\label{2}\ee
describes the holographic dynamics of a $(2+1)$-dimensional theory, with an unbroken U(1) symmetry driven by the most relevant scalar operator, dual to $\phi$.
%We will call it the Einstein-Maxwell-Dilaton$_0$ (EMD$_0$) class of actions.
The general IR asymptotics of these actions were explored in \cite{cgkkm}, while simpler cases with a constant potential have been analyzed earlier, \cite{taylor,KT}.
The analysis of \cite{cgkkm}, supplemented with a further interpretation and study of the IR asymptotics in \cite{gk} indicated the existence of large classes of quantum critical
extremal solutions at finite charge density, characterized by a Lifshitz exponent $z$ and a hyperscaling violation exponent $\theta$ (see also \cite{sachdev,dong,Edalati:2012tc} for an analysis of the interesting implications of hyperscaling violation).

The range of applicability of the class of actions in (\ref{1}) is wider than it appears. For example adding more scalars to the theory, the solutions, and in
particular their IR asymptotics can still be described by solutions to (\ref{1}) with $\phi$ being the appropriate linear combination that defines the RG flow.

In a different direction,  the  action $S_M$ in (\ref{1}) can be interpreted as the effective action of a holographic theory with a U(1) symmetry and  a complex charged scalar.
To motivate this interpretation, we start from a U(1) gauge field and a complex scalar $\Psi$ that is charged under the U(1) gauge symmetry, dual to a charged scalar operator.
 The most general two-derivative effective holographic action %(EHA)
of such fields  can be parametrized after suitable field redefinitions as\footnote{We drop again the $F^*F$ term.}
\be
S=M^{2}\int d^{4}x\sqrt{-g}\left[R-{G(|\Psi|)\over 2}|D\Psi|^2+\tilde V(|\Psi|)-{\tilde Z(|\Psi|)\over 4}F^2\right]
\label{3}
\ee
with the standard covariant derivative as
\be
D_{\m}\Psi=\partial_{\m}\Psi+iq A_{\m}\Psi\,.
\label{4}\ee
The functions $G,\tilde V, \tilde Z$ are a priori arbitrary, and they capture the appropriate effective dynamics.
There are two possibilities in this context :
\begin{enumerate}
\item
The phase of unbroken symmetry where $\Psi=0$.
In that case the saddle points of the system are determined from the simple Einstein-Maxwell-AdS  action
 \be
 S_0=M^{2}\int d^{4}x\sqrt{-g}\left[R-{Z(0)\over 4}F^2+V(0)\right]
\label{6}\ee
that has been studied extensively over the last few years.

\item
The phases with broken  U(1) symmetry.
This will be spontaneous symmetry breaking if the  boundary source is zero, or explicit symmetry breaking if the source is non-zero, \cite{Gubser:2008px}.
In the former case, this generates superfluidity, \cite{Gubser:2008px,HoloSc}.\footnote{And superconductivity if the global symmetry is weakly gauged.}
Assuming solutions with non-trivial $\Psi$ and  changing variables to $\Psi=\chi e^{i\theta}$, the action becomes
\be
S=M^{2}\int d^{4}x\sqrt{-g}\left[R-{G(\chi)\over 2}\left[(\pa_{\m} \chi)^2+\chi^2(\pa_{\m}\theta+qA_{\m})^2\right]+\tilde V(\chi)-{\tilde Z(\chi)\over 4}F^2\right]
\label{5}\ee
At finite density, the electric potential $A_t$ is non-trivial.
We may choose the gauge $\theta=0$ and  change variables $\chi\to \phi$ so that the kinetic term of $\phi$ is properly normalized
to finally obtain the action (\ref{1}). The three arbitrary functions here ($G,\tilde V,\tilde Z$) translate to the three arbitrary functions in (\ref{1}) (with $X=0$).
Therefore, (\ref{1})  with $W(\phi)\not=0$ describes holographic physics in the U(1) symmetry-broken phase.
Superfluidity in such models with generalized couplings has been considered in a number of works, \cite{Gauntlett:2009,EMDmassive}. Note also that the scalar $\theta$ is the source of the Goldstone boson of the broken U(1) symmetry if the breaking is spontaneous.

\end{enumerate}

\subsection{A classification of quantum critical points in phases with (un)broken symmetry}

In this paper we will describe the landscape of quantum critical points of theories described by the actions (\ref{1}) and (\ref{2}). The first criterion one may consider is the scaling symmetries of the IR geometry, and their relation to the IR behavior of the scalar field.

\paragraph{Scaling symmetries of holographic quantum critical points\\}
The metrics we will study take the generic form, \cite{gk}
\be \label{hyperscalingviolating}
\ud s^2=r^\theta\left(-\frac{\ud t^2}{r^{2z}}+\frac{L^2\ud r^2+\ud x^2+\ud y^2}{r^2}\right)
\ee
and are extremal (they have zero temperature). They display both a dynamical exponent $z$ and a hyperscaling violation exponent $\theta$, so that at finite temperature, the thermal entropy scales like:
\be
S\sim T^{\frac{2-\theta}{z}}\,.
\ee
The dual field theory has effectively $d_{eff}=2-\theta$ spatial dimensions at small temperatures, \cite{sachdev}. Hyperscaling violation $\theta\neq0$ can be engineered by letting the scalar run logarithmically, while letting it settle to a constant leads to $\theta=0$. More properties of these metrics are detailed in Appendix \ref{app:hyper}.

An important ingredient in a theory with a U(1) gauge boson is the scaling of the gauge field. This is given in general by
\be
A_t=Q~r^{\zeta-z}
\label{zeta}\ee
with $\zeta$ being a novel critical exponent controlling the impact of the charge density on the physics of the critical theory.\footnote{We would like to thank Sean Hartnoll and Liza Huijse for discussions on this point.}
In the most generic case  (see section \ref{section:massiverunning}), $\zeta$ is defined independently from the other exponents $z$ and $\theta$: when hyperscaling is intact ($\theta=0$), then generically we still have $\zeta\neq0$.
The triplet of critical exponents $(\theta,z,\zeta)$ is characterising QC theories at  non-zero charge density. In the presence of several U(1) symmetries
there is a non-trivial exponent $\zeta$ for each of the associated charge densities.

Hyperscaling violation is intimately linked with the running of the scalar field in the IR and therefore one may classify phases by considering the effective scalar potential derived from \eqref{1}
\be \label{EffPot}
	\Box\phi+\frac{\ud V_{eff}(\phi)}{\ud\phi}=0\,,\qquad V_{eff}(\phi)=V(\phi)-\frac{Z(\phi)}{4}F^2-\frac{W(\phi)}2A^2\,.
\ee
Two qualitative behaviors will be possible.

\subparagraph{{\em The scalar field settles down to a finite constant $\phi_\star$ in the IR, which extremizes $V_{eff}$:}}
\be \label{ScalExtr}
	\left.V'_{eff}\right|_{\star}=\left.\frac{\ud V_{eff}(\phi)}{\ud\phi}\right|_{\phi=\phi_{\star}}=0
\ee
This expression will generically be valid both in the unbroken ($W(\phi)=0$ identically) and broken phase ($W(\phi)\neq0$, though $W_\star=W(\phi_\star)=0$ at the extremum is allowed). The IR fixed points will all be hyperscale-invariant.

In the unbroken phase, both AdS$_4$ (section \ref{section:AdS4unbr}, charge$\to 0$) and AdS$_2\times R^2$ (section \ref{section:AdS2xR2unbr}, charge remains finite) fixed points can occur in the IR. In the broken phase, if the current dual to the gauge field in the bulk is irrelevant\footnote{We define the current to be irrelevant in the IR, when the stress tensor of the gauge field in the IR is negligible compared with the Einstein tensor.} the IR geometry is still an AdS$_4$ domain wall (section \ref{section:neutralDWconstant});  however, if the current is relevant, the asymptotics become Lifshitz ($W_\star\neq0$, section \ref{section:Lifshitz}) or AdS$_2\times R^2$ ($W_\star=0$, section \ref{section:AdS2xR2}).

\subparagraph{{\em The scalar field runs in the IR asymptoting to $\pm\infty$:}\\}
This can happen both in the unbroken  or broken phases, and one will have hyperscaling violating phases ($\theta\neq0$) which may preserve Poincar\'e invariance ($z=1$) or break it ($z\neq1$). The former case will arise for IR phases described by neutral dilatonic domain walls, the latter when charge backreacts on the IR metric. To characterize such phases, we will find convenient to adopt exponential asymptotics for the scalar coupling functions
\be \label{EMDmassiveCouplings}
	Z(\phi)\underset{IR}{\sim}Z_0 e^{\ga\phi},\qquad V(\phi)\underset{IR}{\sim}V_0 e^{-\delta\phi},\qquad W(\phi)\underset{IR}{\sim}W_0 e^{\epsilon\phi}\,.
\ee
 As we explain below, this is without loss of generality, in almost all cases. Such asymptotics are also well-motivated from the point of view of low-energy supergravity theories.

The unbroken symmetry case described by (\ref{2}) was analyzed  in detail in \cite{cgkkm} and all quantum critical solutions with running scalars were found (sections \ref{section:neutralDWunbr} and \ref{section:masslessrunning}). If the symmetry is broken, hyperscaling violating solutions also exist, and can be neutral (section \ref{section:neutralDWrunning}) or charged (sections \ref{section:massiverunning} and \ref{section:masslessrunningbr}). At leading order, two classes are similar to the running solutions with unbroken symmetry (sections \ref{section:masslessrunningbr} and \ref{section:neutralDWrunning}), while a third displays symmetry-breaking even at leading order (section \ref{section:massiverunning}).

Moreover, even when the asymptotic behavior of such functions is not dictated by a pure exponential, the exponential solutions still give the leading asymptotic behavior.
More precisely if we define
\be
\gamma={\rm Infimum}\{\gamma_0\in R:  { \lim_{\phi\to\infty}e^{-\gamma_0\phi}Z(\phi)>0}\}
\ee
\be
\e={\rm Infimum}\{\e_0\in R:  {\lim_{\phi\to\infty}e^{-\e_0\phi}W(\phi)>0}\}
\ee
\be
\delta={\rm Infimum}\{\delta_0\in R:  { \lim_{\phi\to \infty}e^{\delta_0\phi}V(\phi)>0}\}
\ee
then the leading order result in the IR (with $\phi\to\infty$) will be given by the functions (\ref{EMDmassiveCouplings})
 Moreover, if $V\sim e^{-\delta\phi}\phi^{2a}$ as $\phi\to\infty$ then the leading behavior of the solution is determined by $\delta$ whereas $a$ determines subleading corrections, \cite{ihqcd,gkmn,cgkkm}.
If any function increases faster than exponential then typically the Gubser bound is violated.

The only exception to the general statements  above are special values of the exponents that lie at the boundary between stable and unstable near-extremal black hole solutions. For the zero density cases, this value is $|\delta|=1$. At finite density and unbroken symmetry the special values satisfy
\be
4+\gamma ^2+2 \gamma  \delta -3 \delta ^2=0\,.
\ee
It is not yet known what is the similar condition in the presence of a non-trivial $W$ function.

 Hyperscaling, Lifshitz solutions ($\theta=0$) can be found when $\d=0$ (flat potential), see sections  \ref{section:masslessrunningbr} and \ref{section:massiverunning}. In section \ref{app:lifts}, we shall also reinterpret the scaling properties of such hyperscaling violating solutions in terms of generalized dimensional reduction, which involves the analytic continuation of the number of reduced dimensions, and see how they may be connected to hyperscaling solutions, \cite{Kanitscheider:2009as,gk,Gouteraux:2011qh}.

\paragraph{IR behavior of scalar coupling functions\\}
It is also useful to consider the behavior of appropriate ratios of the coupling functions $V(\phi)$, $Z(\phi)$ and $W(\phi)$ in the IR, as the scalar $\phi$ tends to a finite value $\phi_\star$ or diverges $\pm\infty$.
There are two distinct cases:\footnote{A third case would be $\lim_{\phi\to\infty}{W(\phi)\over V(\phi)Z(\phi)}\to \infty$, but this does not correspond to any consistent scaling solution in our analysis, as this limit is incompatible with the equations of motion.}
\begin{enumerate}

\item $\lim_{\phi\to\infty}{W(\phi)\over V(\phi)Z(\phi)}=$ finite. In such a case we have a novel class of scaling solutions that are generalized Lifshitz solutions characterized by a Lifshitz, hyperscaling violation and charge exponents ($z,\theta,\zeta$), see section \ref{section:massiverunning}.

\item $\lim_{\phi\to\infty}{W(\phi)\over V(\phi)Z(\phi)}=0$. In this case  the leading IR solutions are the scaling solutions of the symmetry preserving phases found in \cite{cgkkm}, corrected by a subleading power series, see sections \ref{section:neutralDWrunning} and \ref{section:masslessrunningbr}.

\end{enumerate}

\paragraph{Fractionalized vs cohesive phases:\\}
Another sharp distinction between quantum critical points can be made by exposing the respective contributions of bulk degrees of freedom inside or outside the extremal ``horizon'' to the boundary charge density, \cite{Iqbal:2011in,Hartnoll:2011fn}. The degrees of freedom behind the horizon correspond to deconfined or fractionalized states. Due to the large $N$ limit,
 common to holographic setups, there are parametrically more ($\mathcal O(N^2)$) degrees of
 freedom hidden behind the horizon than outside it.
 The degrees of freedom ``behind" the extremal horizon  generate a non-zero electric flux in the deep IR
\be \label{ElFlux}
	\frac1{4\pi}\int_{\mathbf R_2}Z(\phi)\star F=-\frac{\omega_{(2)}}{4\pi}Z(\phi)\frac{C(r)A_t'(r)}{\sqrt{Br)D(r)}}\neq 0\,,
\ee
where $\omega_{(2)}$ is the volume of the spatial section of the metric, and we have used \eqref{EMDmassiveCouplings} for the gauge coupling in the IR as well as the Ansatz \eqref{8} for the metric. The phases where the electric flux does not vanish in the IR have been named \emph{fractionalized}, in reference to the fact that they are dual to gauge-variant operators (also charged under the U(1)) in ``deconfined'' phases (such as quarks). The charged degrees of freedom outside the extremal horizon do not source any electric flux in the deep IR and are dual to ``confined'' gauge-invariant operators: they realize \emph{cohesive} phases.

By considering both exact and power series solutions, we are able to accommodate cohesive and fractionalized phases both, with a single scalar field. The power series behaviors are obtained by requiring terms proportional to $V_0$, $Z_0$ or $W_0$ to become subleading in the field equations. Then the solution will be expressed as a leading order piece, which is an exact solution of (a truncation of) the equations of motion, supplemented by a subleading power series. The detailed analysis is presented in appendix \ref{conv} for completeness.

\paragraph{Quantum critical points and lines\\}

An interesting finding of the present work is that quantum critical geometries are generic in the IR, be they scale invariant or hyperscaling violating.
Moreover this is independent of whether we are in U(1)-preserving/violating, cohesive/fractionalized phases.

\begin{figure}
\centering
\includegraphics[width=\textwidth,trim=0cm 3.5cm 0cm 2cm, clip=true]{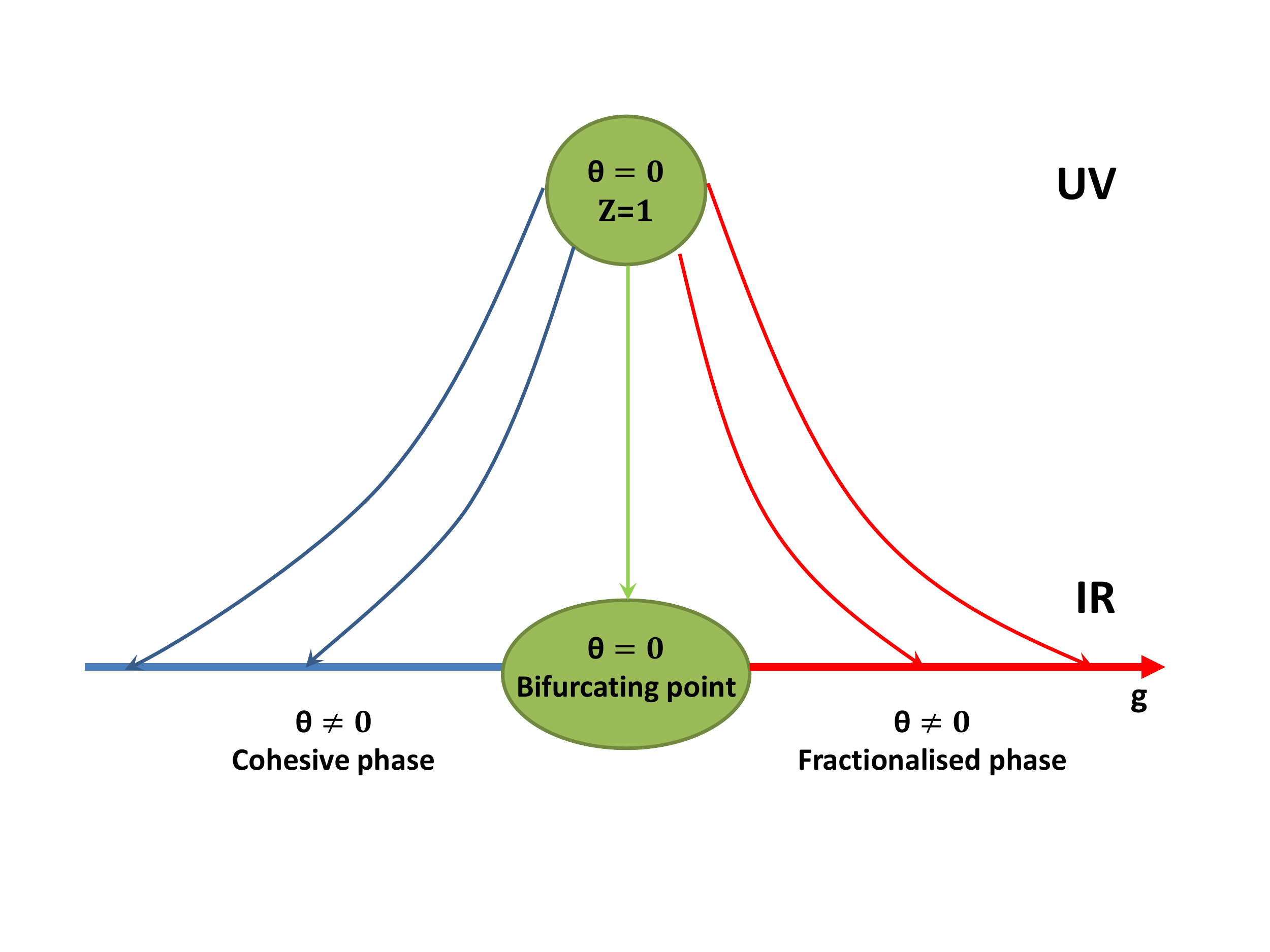}
\caption{Depiction of a schematic RG flow at zero temperature with a bifurcation point (see also \cite{Donos:2012js}). The flow interpolates between AdS$_4$ in the UV and, in the IR, either an unstable, scale invariant ($\theta=0$) critical point, or two hyperscaling violating ($\theta\neq0$) critical lines on each side.}
\label{Fig1}
\end{figure}

This distinction between (partially) fractionalized and cohesive phases has been tied to a specific kind of quantum phase transitions describing the onset of fractionalization, \cite{Hartnoll:2011pp, Hartnoll:2012pp,Adam:2012mw}. They involve a scale invariant IR quantum critical point, which sits at a bifurcation in the holographic RG flow (that is, it has both a relevant and an irrelevant perturbation), see Fig. \ref{Fig1}. As it is unstable, this IR fixed point can only be reached from the UV boundary at the price of fine-tuning the sole dimensionless coupling of the boundary theory, $g=g_{\mathcal O}/\mu$, where $\mu$ is the chemical and $g_{\mathcal O}$ the coupling for the operator $\mathcal O$ dual to the bulk scalar field responsible for driving the IR asymptotics.

Away from this critical value, the relevant deformation is picked up and drives the flow away. Mediated by this unstable critical point (Lifshitz  $z\neq1$, \cite{Hartnoll:2011pp} or AdS$_4$ $z=1$, \cite{ Hartnoll:2012pp,Adam:2012mw}), a phase transition occurs between a cohesive and a fractionalized phase, described by hyperscaling violating geometries. In \cite{ Hartnoll:2012pp,Adam:2012mw}, an auxiliary neutral scalar was responsible for supporting the IR geometry \eqref{hyperscalingviolating}. This meant that, in order to allow for cohesive phases, the dual current needed to be irrelevant and did not backreact on the leading order IR geometries. We will find solutions where the dual current is relevant in the IR and still permits a cohesive phase.

These works gave very interesting insights and we would like to make the picture more precise. Scale invariant solutions correspond to isolated quantum critical \emph{points} in a line where $g$ is varied. There can exist a relevant and an irrelevant perturbation around this point, and then the value of $g$ has to be fine-tuned for the RG flow to reach this point. For generic values of $g$, the flow picks up the relevant deformation lands into a quantum critical \emph{line}, which is a continuous collection of hyperscaling violating, stable quantum critical points. The reason why hyperscaling violating solutions can be reached for continuous values of $g$ can be traced back to the exponential behavior of the scalar couplings, \eqref{EMDmassiveCouplings}.  For logarithmically running solutions, one of the deformations collapses to zero and actually corresponds to a new scaling symmetry: a constant shift of the scalar can be absorbed in a redefinition of the charge, \cite{KT}. Using this scaling symmetry, $g$ can be varied continuously and a quantum critical line develops.

The impression that we have a continuous line of hyperscaling-violating critical points is an illusion to leading order in $1/N_c$.
The scaling symmetry that controls the fixed line is also rescaling the hyperscaling-violating scale $\ell$, characteristic of these critical points, and therefore affects no physical quantities.  As many of these solutions can be lifted to scale invariant solutions (see section \ref{app:lifts}), this scaling symmetry is naturally reinterpreted as a scaling of the volume of the higher-dimensional transverse space.

Despite this, we expect that to the next order in $1/N_c$, they will develop dependence on a new parameter. This is more easily seen by realizing that tree level string theory is independent of volume (as also leading order large-$N_c$ theories). However, at one string loop, a new dimensionless parameter, ${\ell/ \ell_s} $, enters the physics\footnote{ $\ell_s$ is the string scale.} and the fixed point expands into a fixed line.

In the main text, we will discuss whether the scale invariant fixed points have one or two irrelevant, zero-temperature deformations. In the first case only will they be unstable and able to mediate a fractionalization transition between hyperscaling violating lines. We shall see that this relevant deformation can be tied either to the gauge field or to the scalar. We will also study whether these lines are stable (sometimes, depending on the parameters, an irrelevant mode might become relevant, or even complex, in which case we expect a dynamical instability). A more complete analysis is presented in appendix \ref{app:LinPert}.

It is also interesting that a related tale has been argued in \cite{kkp} which has successfully compared holographic transport  to recent very low-temperature data on cuprate superconductors. Indeed this comparison suggests that the $T\to 0$ limit of cuprates,  in the overdoped phase,  may be described by a quantum critical line, \cite{kkp}.

\section{Quantum Criticality in symmetry-preserving phases}

In such a context the relevant effective action is (\ref{2}), which has been analyzed in the past, \cite{cgkkm,gk}. We will describe first  zero, then finite density phases.

\subsection{Quantum Criticality at zero density: cohesive phases}

At zero charge density and chemical potential,  the gauge potential $A_t$ is zero.\footnote{In the unbroken case, $W=0$, at zero charge density, a constant $A_t$
(identified with the chemical potential) can be tolerated and solves the ED equations of motion with zero charge density.
In  a broken phase, $W\not=0$, a non-zero constant piece always induces a non-trivial charge density.}
The relevant action for such solutions is therefore the Einstein-Dilaton (ED) action, namely (\ref{2}) with $Z=0$.
Such actions have been holographically analyzed in full generality in several works including \cite{ihqcd,gkmn,fhr}.

We will take the potential to be non-negative.\footnote{Note that we are using for convenience the
 opposite sign from the standard literature. Therefore a positive potential is Anti-deSitter-like.}
 This assumption is motivated by experience in string theory, and this seems a robust conclusion in the absence of orientifold planes
 based on non-go theorems, \cite{manu} and the fact that flux-induced potentials are positive definite.
De Sitter-like regions have been  advocated in the presence of orientifolds, but the controlled construction of such vacua remains to be done.
We will also take the metric to be Lorentz invariant, setting $z=1$ in \eqref{hyperscalingviolating}
\be
\ud s^2=r^{\theta-2}(L^2\ud r^2+\eta_{\mu\nu}\ud x^\mu\ud x^\nu)\,.
\label{7}\ee
Two cases can be distinguished, whether the scalar settles to a constant or runs logarithmically in the IR.

\subsubsection{Constant scalar \label{section:AdS4unbr}}
The conventional fixed points correspond to finite critical points of the potential, $V'(\phi_*)=0$, with $\phi_*$ finite and by assumption $V(\phi_*)>0$.\footnote{The case
 $V'(\phi_*)=0$ and $V(\phi_*)=0$ is degenerate and is part of the second item.}
The scale-invariant saddle-point solution corresponds to an $AdS_4$ metric and constant scalar $\phi=\phi_*$:
\be \label{AdSDW2}
\ud s^2=\frac1{r^2}\left(-\ud t^2+\ud x^2+\ud y^2 +L^2\ud r^2\right),\quad L^2=\frac{6}{V_\star}\,.
\ee

If $V''(\phi_*)>0$ then this corresponds to an IR fixed point with respect to the operator dual $\phi$, and $\phi$ is irrelevant (positive mass$^2$) in that fixed point.

If $V''(\phi_*)<0$ then this corresponds to a UV fixed point with respect to the operator dual $\phi$, and $\phi$ is relevant (negative mass$^2$) in that fixed point.

Finally if $V''(\phi_*)=0$ then the operator is classically marginal, and the nature of the fixed point (UV or IR) is decided by the first non-zero derivative of the potential.

Including a gauge field perturbation in the background \eqref{AdSDW2} (see appendix \ref{app:AdS4}), we find
\be\label{GaugeAdS}
	A=A_1^- + A_1^+ r\,.
\ee
$A_1^-$ is a marginal deformation, while $A_1^+$ is relevant. Sourcing only $A_1^-$ gives a cohesive phase (there is no IR electric flux), while sourcing $A_1^+$ gives a fractionalized phase, at the price of introducing a relevant deformation. According to \eqref{ElFlux}, the  IR electric flux reads:
\be \label{}
	\frac1{4\pi}\int_{\mathbf R_2}Z(\phi)\star F=-\frac{\omega_{(2)}Z_\star}{4\pi L}A_1^+\,.
\ee

\subsubsection{Running scalar\label{section:neutralDWunbr}}
 These are ``singular" points where  $V$ vanishes or diverges. Taking $V$ to behave exponentially: $V\sim V_0~e^{-\delta\phi}$ as $\phi\to \pm \infty$, there are running scaling solutions which violate hyperscaling \eqref{hyperscalingviolating}:
\be \label{neutralDW2}
	\begin{split}
	&\ud s^2=r^\theta\left(\frac{L^2\ud r^2-\ud t^2+\ud x^2}{r^2}\right),\qquad L^2=\frac{(\theta-3 ) (\theta-2 )}{V_0}\,,\\
	&e^\phi = r^{\sqrt{\theta (\theta -2)}}, \qquad \theta=\frac{2\delta^2}{\delta^2-1}\,.
	\end{split}
\ee
Depending whether the IR region is near $r=0$ ($\theta>3\Leftrightarrow1<\d^2<3$) or $r\to\infty$ ($\theta<0\Leftrightarrow\d^2<1$) and the sign of $\d$, $\phi$ diverges to $\pm \infty$ and $V(\phi)$ asymptotes either to $+\infty$ or zero (note that the solution is not defined for $0<\theta<2$, and $r$ is time-like for $2<\theta<3$).\footnote{The NEC \eqref{NEC} only yields the inequality $\theta(\theta-2)>0$, which is not enough to have a well-behaved solution.}
Such solutions are singular in the conventional sense although for $\delta\in (-\sqrt{3},\sqrt{3})$ ($\theta<0$ or $\theta>3$) they satisfy the Gubser bound, \cite{gubser} and are therefore expected to be resolvable singularities, \cite{cgkkm}.

In string theory such solutions may arise in different guises. They may be decompactification points where an internal radius scalar diverges or where a coupling constant becomes strong. In all such cases there is also a resolution of the singularity. In the decompactification case, one should describe them in the higher-dimensional theory (as in \cite{gk}). In the strong coupling case one should include the corrections (as in \cite{kachru-coup}).
As there is a need that such singularities are resolved in the IR, such geometries can be considered as approximate scale invariant theories. Depending on parameters, the geometry describes an almost exact scale invariant regime at an intermediate range of energies before it runs off to the regular resolved geometry regime in the ultimate IR, like in the case of ``walking" QFTs.

Although the CFTs are simple to find in the holographic context, and they fall into the two categories above, the pattern of RG flows
 is a more complex problem. An important question in particular concerns the nature of RG flows that interpolate between CFTs belonging to the first and second class above. Although some progress has been made in special cases in the past \cite{cgkkm,perl} the essential problem remains unsolved.

Including a gauge field perturbation in the background \eqref{neutralDW2} (see appendix \ref{app:NeutralHypVio}), we find
\be\label{GaugeAdS1}
	A=A_1^- + A_1^+ r^{\b^q_+}\,,\quad \b^q_{+}=1-\gamma\sqrt{\theta(\theta-2)}\,.
\ee
$A_1^-$ is a marginal deformation, but the nature of $A_1^+$ depends on the value of $\gamma$ and $\theta$: it is irrelevant only  when $ \b^q_{+}(3-\theta)<0$.
%if $\theta<0$ (the IR is $r\to+\infty$), then it is irrelevant when $\gamma>1/\sqrt{\theta(\theta-2)}$;  if $\theta>2$ (the IR is $r\to0$), then it is irrelevant when $\gamma<1/\sqrt{\theta(\theta-2)}$.
Sourcing only $A_1^-$ gives a cohesive phase (there is no IR electric flux), while sourcing $A_1^+$ gives a fractionalized phase.\footnote{Such a fractionalized IR phase can be obtained in an analytical flow using the solutions described in section 8 of \cite{gk}.} According to \eqref{ElFlux}, the  IR electric flux reads:
\be \label{}
	\frac1{4\pi}\int_{\mathbf R_2}Z(\phi)\star F=-\frac{\omega_{(2)}Z_0}{4\pi L} \b^q_{+}A_1^+\,.
\ee

\subsection{Quantum Criticality at finite density: fractionalized phases}

In a CFT at finite charge density, the conformal invariance is spontaneously broken by the charge background. It came as a surprise that in the simplest holographic description of a finite density state, the RN extremal black  hole (a zero-temperature solution of the action in (\ref{6})), a novel scaling symmetry emerged  in the IR regime, realized by an AdS$_2$ geometry, \cite{mit}. As we show this persists in more general situations.

Using the EHT with action $S_0$ in (\ref{2}) describing the holographic physics in a phase with unbroken U(1) symmetry we can classify all quantum critical geometries at finite temperature, both for a constant or running scalar.

\subsubsection{Constant scalar\label{section:AdS2xR2unbr}}

 Quantum critical geometries with constant and finite $\phi$ are analogous to the standard AdS fixed points of the previous subsection.
Such geometries are $AdS_2\times R^2$ and generalize the analogous geometry of the RN black hole.
The important difference here is that the value  $\phi_*$ of the scalar is no-longer an extremum of $V(\phi)$ but of $V_{eff}(\phi)$ in \eqref{EffPot}, which reads on-shell:
\be
 \left.{\ud V_{eff}(\phi)\over \ud\phi}\right|_{\phi=\phi_*}=V'_\star+\frac{Z'_\star}{Z_*}V_*=0\,.
\label{2.3}\ee

For each such $\phi_*$ the metric and gauge field in the solution are
\be
ds^2={L^2\over r^2}(\ud r^2-\ud t^2)+\ell^2(\ud x^2+\ud y^2)\sp A=\mu-{Q\over r}
\label{9}\ee
with
\be
L^2={1\over V(\phi_*)}\sp Q^2={2\over Z(\phi_*)V(\phi_*)}
\label{10}\ee
while $\ell$ remains undetermined.
$Q$ is the charge (density) in $AdS_2$, and it is fixed by the equations.
If however, this solution is the IR limit of a flow of  an (AdS$_4$ \eqref{AdSDW2} or hyperscaling violating \eqref{neutralDW2}) UV solution then the UV charge density is proportional to $Q/\ell^2$, and in the IR it determines the value of $\ell$. This can be verified explicitly both in the RN extremal solution as well as in the $\gamma=\delta$ solutions in \cite{cgkkm}.\footnote{The $\gamma\delta=1$ solutions in this reference provide another example, though the IR geometry violates hyperscaling.}

From \eqref{2.3} we observe that a scaling solution at zero density (determined by $V'=0$) remains scale invariant if $Z'=0$ at the same value of $\phi$.
 In a class of examples, this happens together with $Z=0$ which in turn implies that the gauge field is absent in the relevant IR CFT, its kinetic term vanishing due to strong coupling. This is indeed what happens to the brane gauge fields when the branes and anti-branes annihilate during tachyon condensation, \cite{yi}.

When $Z'\not=0$, the value of $\phi$ is shifted, or there is no solution to (\ref{2.3}) and the only IR critical points have a runaway scalar, see below.

    The conditions for the presence of an AdS$_2\times R^2$ solution generalize simply in the case of $N\geq 1$ U(1) gauge fields $A^I_{\m}$, and $m\geq 1$ (neutral) scalars, $\phi_i$.
    \be
    V(\phi^i_*)\sum_{I=1}^N~Z_I(\phi_*^i)Q_I^2=2\sp {d \over d\phi_i}\log (V\sum_{I=1}^N~Z_IQ_I^2)\Big|_{\phi^i=\phi_*^i}=0\,.
   \label{11} \ee
    Note that now we have $N$ charge densities and only one combination of them is determined by the conditions above. 
    By varying the others, one can generically find solutions to the minimization conditions (\ref{11}).
    Moreover now, the critical points become critical manifolds as one can vary continuously the ratios of charge densities.

\subsubsection{Running scalar\label{section:masslessrunning}}

 We now turn to quantum critical geometries driven by  the scalar running off to $\pm\infty$. Such geometries were found and classified in \cite{cgkkm,gk}:\footnote{The solutions themselves were known earlier, see \cite{cgs} and the relevant references in \cite{cgkkm}. Their interpretation and physical analysis in the holographic context was done in \cite{cgkkm} and the realization that they represent quantum critical points with (generically) hyperscaling violation was advocated in \cite{gk}.}
   \be \label{MasslessSolEMD}
	\begin{split}
	&\ud s^2=-\left(\frac{r}{\ell}\right)^{\theta-2z}f(r)\ud t^2+\left(\frac{r}{\ell}\right)^{\theta}\frac{L^2\ud r^2}{r^2f(r)}+r^{\theta-2}\ud \vec{x}^2\,,\quad L^2=\frac{(1+z-\theta ) (2+z-\theta )}{V_0}\\
	&f(r)=1-\left(\frac{r}{r_h}\right)^{2+z-\theta},\quad e^\phi=\left(\frac{r}{\ell}\right)^{\frac\theta\d},\quad A=\sqrt{\frac{2(-1+z)}{Z_0(2+z-\theta )}}\left(\frac{\ell}{r}\right)^{2+z-\theta}f(r)\,\ud t,\\
	& \d=\frac{\theta }{\sqrt{(\theta-2)(2-2 z+\theta)}}\,,\qquad  \g=\frac{4-\theta }{\sqrt{(\theta-2)(2-2 z+\theta)}}\,,\\
	&\theta=\frac{4\d}{\g+\d}\,,\qquad z=\frac{4+\gamma ^2+2 \gamma  \delta -3 \delta ^2}{(\gamma -\delta ) (\gamma +\delta )}
\,\zeta=\theta-2
\,.	\end{split}
\ee
The constraints such that this solution is consistent were derived in \cite{cgkkm}, and are more stringent than those imposed by the NEC. The allowed parameter range depending on the location of the IR regime is (see discussion in appendix \ref{app:hyper}):
\be\label{IRconstraintsEMDmassless}
\begin{split}
IR:\,r\to0\,:&\quad \left[2<\theta\leq3\,, \,z<\theta-2\right],\quad \left[\theta>3\,,\,z<1\right], \\
IR:\,r\to+\infty\,:&\quad \left[\theta\leq0\,, \,z>1\right],\quad \left[0<\theta<2\,,\,z>\frac{2+\theta}2\right].
\end{split}
\ee
There is always an IR curvature singularity, except for the range $0<\theta<2$ with the IR at $r\to+\infty$.
$r_h$ is a free integration constant related to the temperature, while the electric charge and chemical potential are
\be
Q=\frac{\omega_{(2)}M^2}{2}\sqrt{\frac{V_0(z-1)}{2(1+z-\theta)}}\ell^{\theta-2}\,,\qquad \mu=-\sqrt{\frac{2(z-1)}{2+z-\theta}}\left(\frac{\ell}{r_h}\right)^{2+z-\theta},
\ee
where $\omega_{(2)}$ is the volume of the spatial directions.
As in the AdS$_2\times R^2$ case above, the charge can be varied by varying the length scale $\ell$. The value of $\ell$ will be connected to UV data.

For special values of the parameters $z$ and $\theta$, there is a larger symmetry, \cite{gk}:
\begin{itemize}
	\item $z\to+\infty$ with $\theta$ finite ($\g=\d$) reduces the metric to AdS$_2\times R^2$ with a constant scalar;
	\item $z,\,\theta\to+\infty$ while keeping their ratio fixed ($\g=-\d$) and changing coordinates to $r^z=\rho$ makes the metric conformal to AdS$_2\times R^2$. This case enjoys special dimensional reduction properties, \cite{gk} (see also section \ref{app:LiftW0=0}) and has been called semi-locally critical: time scales but space does not. Green's functions still depend on momentum, which gives rise to interesting behavior, \cite{semilocal};
	\item $\theta=0$ ($\d=0$) gives a Lifshitz metric, \cite{taylor,KT}.
\end{itemize}

One finds that the electric flux \eqref{ElFlux} generated by \eqref{MasslessSolEMD} is finite in the IR and proportional to $Q$, so the degrees of freedom are  fractionalized.

\section{Quantum Criticality in symmetry-breaking phases}

In this section, we will focus on symmetry-breaking phases described by \eqref{1} and present a classification of all quantum critical points possible in the IR: fractionalized or cohesive (vanishing of electric flux in the IR), neutral or charged (subleading electric potential in the IR), hyperscaling or not (constant or running scalar), with unbroken or broken symmetry  ($W(\phi)=0$ or not). To obtain charged cohesive phases, it is crucial that the charged scalar itself supports the IR geometry \eqref{hyperscalingviolating} (that is, it should not be subleading in the IR).

\subsection{Quantum Criticality at zero density: cohesive phases\label{section:neutralDW}}

 We first consider the cohesive phases, that is phases where the electric flux \eqref{ElFlux} vanishes in the IR, and distinguish once more between scale invariant and hyperscaling violating solutions. Note that in some cases these phases can become (partially) fractionalized, but we shall describe them in this subsection for conciseness.

\subsubsection{Constant scalar\label{section:neutralDWconstant}}

Setting the charge to zero in the field equations \eqref{MaxEqmassive}-\eqref{DilEqMassive}, and requiring the scalar to settle in an extremum of the scalar potential, \eqref{ScalExtr}, one is simply left with AdS in the Poincar\'e patch:
\be \label{AdSDW}
\ud s^2=\frac1{r^2}\left(-\ud t^2+\ud x^2+\ud y^2 +L^2\ud r^2\right),\quad L^2=\frac{6}{V_\star}\,.
\ee
 This domain wall can be used to set symmetry-breaking  boundary conditions both in the UV and in the IR, \cite{Gubser:2008wz,Gauntlett:2009,GubserNellore,Horowitz:2009ij}.

We now turn to the study of the zero temperature deformations around \eqref{AdSDW} (see appendix \ref{app:AdS4} for a full analysis), and wish to determine whether they are irrelevant or relevant in the IR ($r\to+\infty$). One mode comes from the scalar field:
\be
		\Delta\phi=\phi_1 r^{\b^\phi_-},\qquad \b^\phi_-=\frac12\left(3-\sqrt{9-4L^2V_\star''}\right)
\ee
and is real and irrelevant if $V''_{\star}<0$, real and relevant if $0<V''_{\star}<V''_{BF}=9/4L^2$, complex and relevant when the scalar mass goes below the Breitenlohner-Friedman bound, $V''_{\star}>V''_{BF}$. The IR AdS$_4$ geometry is dynamically unstable, as the dimension of the dual operator to $\phi$ becomes complex. It has been shown that a Lifshitz geometry (see section \ref{section:Lifshitz}) could arise instead, \cite{GubserNellore,Horowitz:2009ij}.

The second irrelevant perturbation is provided by turning on the gauge field
\be
 \Delta A=A_1r^{\b^q_-},\qquad	\b^q_-=\frac12\left[1-\sqrt{1+\frac{4L^2 W_\star}{Z_\star}}\right].
\ee
Assuming $Z_\star>0$ and $W_\star>0$, $\b^q_-$ is always real and negative. When the U(1) symmetry is broken in the IR, this phase is always cohesive as the electric flux vanishes in the IR.

When the U(1) symmetry is not broken in the IR ($W_\star=0$), then $\beta^q_-=0$, as one would find for the chemical potential in the UV. One can then consider the conjugate deformation $\b^q_+=1$: it is a relevant deformation, but now generates a constant IR electric flux, describing a (partially) fractionalized phase. According to \eqref{ElFlux}, the  IR electric flux reads:
\be \label{}
	\frac1{4\pi}\int_{\mathbf R_2}Z(\phi)\star F=-\frac{\omega_{(2)}Z_\star}{4\pi L} A_1^+\,.
\ee

%As a conclusion, one needs to impose $V''_{\star}>0$ (negative mass squared) for the AdS$_4$ IR fixed point to have a relevant deformation and be a (dynamically) unstable bifurcation point, while if $V''_{\star}<0$, the two deformations are irrelevant and the fixed point is stable.

As a conclusion, there will be a (dynamically) unstable bifurcation point if  $V''_{\star}>0$ and $W_\star\neq0$, or including the $\beta_q^+$ deformation when $V''_{\star}<0$ and $W_\star=0$. In all other cases, there are two irrelevant deformations and this a stable fixed point.

\subsubsection{Running scalar\label{section:neutralDWrunning}}

Turning to solutions with a running scalar, the neutral solution \eqref{neutralDW2} can still describe the leading order IR solution when the $U(1)$ is broken, if we let the terms proportional to the gauge field and its derivative be subleading in the IR, see appendix \ref{appC.5}. If both are subleading and the mass term in Maxwell's equation as well, we find the power series (with the scalar functions parametrized as in  \eqref{EMDmassiveCouplings}):
\be \label{neutralDWPowerSeries}
	\begin{split}
	&\ud s^2=B(r)\ud r^2-D(r)\ud t^2+r^{\theta-2}\ud x^2\,,\quad L^2=\frac{(\theta-3 ) (\theta-2 )}{V_0}\,,\\
	&D(r)=r^{\theta-2}\left(1+\sum_{n\geq1}\mathfrak d_n r^{n\alpha_2}\right),\quad B(r)=L^2r^{\theta-2}\left(1+\sum_{n\geq1}\mathfrak b_n r^{n\alpha_2}\right),\\
	&e^\phi = r^{\sqrt{\theta (\theta -2)}}\left(1+\sum_{n\geq1} \varphi_n r^{n\alpha_2}\right), \quad A=Q\left(1+\sum_{n\geq1}\mathfrak a_n r^{n\alpha_1}\right),\\
	&\theta=\frac{2\delta^2}{\delta^2-1}\,,\quad \a_1=(\epsilon -\gamma)  \sqrt{(\theta-2)\theta }+\theta\,,\quad	 \a_2=2+\epsilon   \sqrt{(\theta-2)\theta }\,,
	\end{split}
\ee
where $\mathfrak d_n$, $\mathfrak b_n$, $\varphi_n$ and $\mathfrak a_n$ are uniquely determined by the field equations and proportional to $Q^2W_0$. The allowed parameter range is the same as for the solution \eqref{neutralDW2} with unbroken $U(1)$, and differs whether the IR is $r\to+\infty$ ($\theta<0$) or $r\to0$ ($\theta>3$).
In order for the power series to be consistent, we should also require that the powers $\alpha_{1,2}$ are subleading in the IR, which translates as the following inequalities on $\epsilon$ and $\gamma$, depending on the sign of $(3-\theta)$:
\be\label{ConstraintsDWSeries1}
	(3-\theta )\left(\theta +(\epsilon -\gamma )\sqrt{\theta (\theta -2)}\right)<0\,,\quad(3-\theta )\left(2+\epsilon \sqrt{\theta (\theta -2}\right)<0\,.
\ee
Turning to linearized deformations around (the leading order part of) \eqref{neutralDWPowerSeries}, two irrelevant deformations are found, driving zero-temperature flows ($\mathcal Q=0$). Both are now zero modes, $\beta^\phi_-=0$, $\beta^q_-=0$, and are respectively a shift of the scalar $\Delta\phi=\phi_1^-$ and of the gauge field charge $\Delta A=A_1^-$ (since $W_0$ is subleading, in contrast to \eqref{eq3.7} below).

 As discussed in section \ref{section:neutralDWunbr}, if we turn on the deformation conjugate to $\beta^q_-$, that is $\beta^q_+=1-\gamma/\sqrt{\theta(\theta-2)}$, this allows for a (partially) fractionalized phase. This is an irrelevant deformation only in the more constrained parameter phase (compared to \eqref{ConstraintsDWSeries1}:
\be\label{ConstraintsDWSeries2}
 (3-\theta )\beta^q_+<0\,,\quad(3-\theta )\left(2+\epsilon \sqrt{\theta (\theta -2}\right)<0\,.
\ee
According to \eqref{ElFlux}, the IR electric flux then reads:
\be \label{}
	\frac1{4\pi}\int_{\mathbf R_2}Z(\phi)\star F=-\frac{\omega_{(2)}Z_0}{4\pi L} \b^q_{+}A_1^+\,.
\ee

Another option is if $Q=0$, in which case the power series in \eqref{neutralDWPowerSeries} vanishes and the solution becomes an exact solution of the field equations (identical to \eqref{neutralDW2}). This means that the mass term in Maxwell's equation is no longer subleading compared to the kinetic term, which will enforce $\epsilon=\gamma-\delta$ in the gauge field fluctuations. The linearized deformations around that solution now read:
\be
	\Delta\phi=\phi_1\,,\qquad \Delta g_{\mu\nu}=\delta\phi_1L^2 r^{\theta-2}\delta_\mu^r\delta_\nu^r
\ee
which is simply a constant shift of the scalar, while the other turns on charge
\be \label{eq3.7}
	\Delta A=A_1^\pm r^{\b^q_\pm},\qquad \beta^q_\pm=\frac12\left(1-\sqrt{\theta(\theta-2)}\g\pm\sqrt{\left(1-\sqrt{\theta(\theta-2)}\g\right)^2+4L^2\frac{W_0}{Z_0}}\right),
\ee
Here it is necessary that $\e=\g-\d$.
Comparing with the previous section \ref{section:neutralDW}, the irrelevant scalar mode has collapsed to zero because of the choice of exponential potential, while for the gauge field deformation setting $\g=\d$ and $\theta=-n$ using the lift in section \ref{app:LiftW0neq0} recovers the result in AdS$_{n+4}$. If the IR is $r\to0$ ($\theta>3$), we should choose $\b^q_+$, otherwise $\b^q_-$  ($r\to+\infty$, $\theta<0$). In their regime of validity, both deformations are always irrelevant irrespective of the value of $\gamma$.
Moreover, we enforce that terms proportional to the gauge field in the scalar equation are subleading with respect to to the potential. This gives the inequality
\be
	\g\sqrt{\theta(\theta-2)}+2-\theta+2\b^q_-<0\Leftrightarrow 2(3-\d^2)\frac{W_0}{V_0Z_0}+(1-\g\d)(\d^2-\g\d-2)>0\,,
\ee
constraining allowed values for $\g$.

As we found two irrelevant perturbations for each case, spontaneous symmetry breaking in the boundary theory can be engineered.
Since the gauge field is subleading in the deep IR, the electric flux \eqref{ElFlux} vanishes, this will always describe a cohesive phase.

\subsection{Quantum Criticality at finite density: Fractionalized phases}

In fractionalized phases, the electric flux \eqref{ElFlux} does not vanish in the IR. The UV charge density is the sum of this flux and any potential bulk charged matter flux.  This can be achieved with scale invariant (section \ref{section:AdS2xR2}) or hyperscaling violating (section  \ref{section:masslessrunningbr}) solutions. The scalar will be either settling in a minimum of the effective potential \eqref{EffPot} or will diverge logarithmically.

\subsubsection{Constant scalar \label{section:AdS2xR2}}
We first start by searching for a scale invariant solution with a constant scalar $\phi=\phi_\star$. This value is determined by extremising the effective potential \eqref{EffPot}.
An AdS$_2\times R^2$ solution ($z\to\infty$) exists provided one sets $W_\star=0$ and
\be
W'_\star= -Z_\star V'_\star-V_\star Z'_\star
\ee
at $\phi=\phi_\star$. Then
\be \label{AdS2xR2br}
\ud s^2=\frac1{r^2}\left(-\ud t^2+\frac{\ud r^2}{V_\star}\right)+\ell^2\left(\ud x^2+\ud y^2\right)\,,\quad A=\sqrt{\frac2{Z_\star}}\frac1r\ud t\,.
\ee
Note that here the functions  $V$, $Z$ and $W$ are a priori arbitrary.  It is easily verified that the electric flux in the IR \eqref{ElFlux} is non-zero:
\be
\frac1{4\pi}\int_{\mathbf R_2}Z(\phi)\star F=\frac{\omega_{(2)}\ell^2}{4\pi}\sqrt{2Z_\star V_\star}\,,
\ee
so this is indeed a fractionalized phase

The linearized perturbations around the solution are detailed in appendix \ref{app:AdS2R2}. One finds four conjugate modes: if they are real, two of them are positive and therefore relevant deformations, while the other two can be negative and irrelevant deformations. If they are complex, then their real part is necessarily positive and equal to $1/2$, and the fixed point is dynamically unstable. A last option is to have only one irrelevant mode, which allows to realize our bifurcation scenario.

We pick $W'_\star=0$ to make the formulae simpler and find:
\be \label{2.6}
\b^1_-=-1\,,\quad \b^1_+=2\,,\quad \b^2_{\pm}=\frac12\left(1\pm\sqrt{1-4\lambda}\right), \quad \lambda=-\frac{2V'_\star{}^2}{V_\star{}^2}+\frac{V''_\star}{V_\star}+\frac{W''_\star}{V_\star Z_\star}+\frac{Z''_\star}{Z_\star}.
\ee
This case encompasses in particular the well studied theory of the charged complex scalar with a quadratic potential, \cite{HoloSc,GubserNellore,Horowitz:2009ij}. We may distinguish three cases. 

If $\lambda<0$, then $\b^2_-<0$, there are two irrelevant perturbations in the IR and one of them can be tuned so that the source of the dual scalar operator is switched off in the UV, leading to spontaneous symmetry breaking and a stable IR fixed point. It would be interesting to investigate whether AdS$_2\times R^2$ could be the IR endpoint of a flow with broken symmetry in the UV, when $W_\star'\neq0$ but $W(0)=0$ (so that AdS$_2\times R^2$ is a solution of the field equations both with and without the scalar turned on).

If $0<\lambda<1/4$, then $\b^2_->0$ and there is only one irrelevant mode, $\b_-^1$. If $\lambda>1/4$ (which corresponds to the effective Breitenlohner-Freedman bound), then $\b^2_-$ is complex with a positive real part.  This happens for instance when the square root in \eqref{2.6} becomes negative, for a large enough scalar charge $W_\star''$ and a small enough scalar mass $V_\star''$. In that case, one also expects an instability (of the phase with broken symmetry or not), and the (effective) Breitenlohner-Freedman bound in AdS$_2$ is violated.

To summarize, this fixed point can mediate a fractionalization transition (in the case $W'_\star=0$) only if $\lambda<0$ (so that there is one irrelevant and one relevant deformation), otherwise it is stable.

\subsubsection{Running scalar\label{section:masslessrunningbr}}

 We now consider the cases where the scalar does not settle in an extremum of the effective potential \eqref{EffPot}, but is allowed to diverge  logarithmically in the IR. Assuming  the asymptotic scalar functions  take the form \eqref{EMDmassiveCouplings}, and $W(\phi)\to0$ at leading order, the field equations \eqref{MaxEqmassive}-\eqref{DilEqMassive} admit the following hyperscaling violating series solution:
\be \label{MasslessSolEMDPower}
	\begin{split}
	&\ud s^2=-D(r)\ud t^2+L^2B(r)\ud r^2+\frac{\ud \vec{x}^2}{r^{2-\theta}}\,,\quad L^2=\frac{(1+z-\theta ) (2+z-\theta )}{V_0}\\
	&D(r)=r^{\theta-2z}\left(1+\sum_{n=1}\mathfrak d_nr^{n\a}\right),\quad B(r)=r^{\theta-2}\left(1+\sum_{n=1}\mathfrak b_nr^{n\a}\right),\\
	&e^\phi = r^{\sqrt{(\theta-2)(2-2z+\theta)}}\left(1+\sum_{n=1}\varphi_nr^{n\a}\right),\quad A=\frac{Q\ud t}{r^{2+z-\theta}}\left(1+\sum_{n=1}\mathfrak a_nr^{n\a}\right),\\
	& Q^2=\frac{2(-1+z)}{Z_0(2+z-\theta )}\,,\quad \a=\frac{4 (-\gamma +\delta +\epsilon )}{\gamma +\delta }\,,\quad \e=\frac{4+\a -2 \theta }{\sqrt{(-2+\theta ) (2-2 z+\theta )}}\,,\\
	& \d=\frac{\theta }{\sqrt{(\theta-2)(2-2 z+\theta)}}\,,\quad  \g=\frac{4-\theta }{\sqrt{(\theta-2)(2-2 z+\theta)}}\,,\\
	&\theta=\frac{4\d}{\g+\d}\,,\quad z=\frac{4+\gamma ^2+2 \gamma  \delta -3 \delta ^2}{(\gamma -\delta ) (\gamma +\delta )}\,.
	\end{split}
\ee
The leading order solution ($n=0$) is  the solution studied in \cite{cgkkm,gk}, see \eqref{MasslessSolEMD},
 while the subleading series coefficients are uniquely determined from the equations of motion and are proportional to the ratio $W_0/(V_0Z_0)$. The constraints on $\theta$ and $z$ are as in \eqref{IRconstraintsEMDmassless}. Note that the exponent $\e$ in W is arbitrary, but should obey the inequality
\be
\frac{4 (-\gamma +\delta +\epsilon )}{\gamma +\delta }(2+z-\theta)>0
\ee
so that the subleading terms in the series are subleading in the IR.\footnote{Note that if $\frac{4 (-\gamma +\delta +\epsilon )}{\gamma +\delta }(2+z-\theta)<0$
there is no IR scaling solution.}
 
Restricting to $\theta\neq1+z$ ensures $V_0\neq0$, and we shall require $V_0>0$. Constraints on the background solution are that $r$ should be spacelike, and that $Q^2>0$, \cite{cgkkm}. These are strictly equivalent to the constraints that can be derived on the metric \eqref{MasslessSolEMD} from the NEC \eqref{NEC}, \cite{dong}.

One finds that the electric flux \eqref{ElFlux} in \eqref{MasslessSolEMD} is finite in the IR:
\be
\frac1{4\pi}\int_{\mathbf R_2}Z(\phi)\star F=\frac{\omega_{(2)}}{4\pi}\sqrt{\frac{2(z-1)V_0Z_0}{1+z-\theta}}
\ee
so the degrees of freedom are (at least partially) fractionalized.

Perturbing \eqref{MasslessSolEMDPower} at the linearized level, we find two irrelevant, zero temperature deformations. The first is a constant mode, which either shifts the scalar by a constant or reparametrizes time. The other one is non-universal and reads
\be
	\Delta \phi=\phi_1 r^{\b_\pm},\quad \b_\pm=\frac12\left(2+z-\theta\pm\sqrt{\frac{2+z-\theta}{2z-2-\theta}\left(18 z^2+2 z-20+16 \theta -19 z \theta +\theta ^2\right)}\right).
\ee
It sources the metric and the gauge field as well, see appendix \ref{app:ChargedFracHypVio}. Depending on where the IR is located ($r\to0,+\infty$), we should select either $\b_+$ or $\b_-$, respectively. Within the allowed parameter range \eqref{IRconstraintsEMDmassless}, the appropriate mode is always an irrelevant deformation. It can never become complex, so there is no dynamical instability associated to the geometry \eqref{MasslessSolEMDPower} in our setup.

\subsection{Quantum Criticality at finite density: Cohesive phases\label{section:massiveSolRunningScalar}}

 We consider now  cohesive phases (the electric flux \eqref{ElFlux} vanishes in the IR) where the gauge field participates at leading order in the IR asymptotics. The scalar will be either settling in a minimum of the effective potential \eqref{EffPot} or diverge  logarithmically. As $W(\phi)\neq0$, the UV theory will be in a symmetry-breaking phase.

\subsubsection{Constant scalar \label{section:Lifshitz}}
 The first possibility is for the scalar to settle in an extremum of the effective potential \eqref{EffPot}, which gives rise to a scale invariant Lifshitz solution, \cite{GubserNellore,Horowitz:2009ij}:
\be
	\begin{split}
	&\ud s^2=-\frac{\ud t^2}{r^{2z}}+\frac{L^2\ud r^2+\ell^2\left(\ud x^2+\ud y^2\right)}{r^2}\,,\qquad L^2=\frac{4+z+z^2}{V_\star}\,,\\
	&A=Qr^{-z}\ud t\,,\qquad Q^2=\frac{2(z-1)}{z Z_\star},\\
	&W_\star=\frac{4(z-1)}{L^2Q^2}\,,\qquad 0=\left.\left[\log\left(VW^{\frac{2(z-1)}{4+z+z^2}}Z^{\frac{z(z-1)}{4+z+z^2}}\right)\right]'\right|_{\phi=\phi_\star}\,.
	\end{split}
	\label{MassiveSolConstantPhi}
\ee
Since we are interested in $z$ finite, we can restrict to $W_\star\neq0$ (see section \ref{section:AdS2xR2} otherwise). Note also that $W_\star>0$ implies $z>0$, which, once combined with requesting the charge $Q$ to be real, yields $z>1$. This is identical to the constraint derived from the NEC \eqref{NEC}. For the specific value $z=2$, this solution can be lifted to a $z=0$ Schr\"odinger solution in an axion-dilaton theory (albeit with a constant dilaton), \cite{LifshitzFromAdS}.

As expected, the electric flux \eqref{ElFlux} vanishes in the IR $r\to+\infty$ like $r^{-2}$, which is in accord with the interpretation that this solution describes a fully cohesive phase, \cite{Hartnoll:2011fn}.

Turning to linearized deformations (see appendix \ref{app:Lifshitz} for details), we find two conjugate pairs of modes,  two of which will typically have negative real parts. Their exact expressions are quite involved. There is always one independent amplitude, setting the strength of the deformation. Selecting a particular model helps the analysis. In \cite{GubserNellore}, the quadratic and quartic potential models were studied, and it was shown that two of the modes could be real and irrelevant in a large portion of the parameter space, giving rise to superfluidity.

To give another example, we set $V'_\star=Z'_\star=W'_\star$ (so that the scalar extremizes separately all individual contributions to its effective potential). The expressions for the modes become:
\be
\begin{split}
	&\b^1_{\pm}=\frac12\left(z+2\pm\sqrt{20-20z+9z^2}\right), \\
          & \b^2_{\pm}=\frac{z+2}2\left(1\pm\sqrt{1-4\lambda}\right),\qquad \lambda=\frac{L^2V''_\star+\frac12L^2Q^2W''_\star+\frac12z^2Q^2Z''_\star}{(z+2)^2}\,.
\end{split}
\ee
 We note that $\b_-^1<0$ if and only if $z<1$ or $z>2$, where only the latter is compatible with the constraints on the metric. If $1<z<2$, then both $\b_\pm^1$ are relevant deformations. On the other hand, $\b_+^2$ is always relevant, while $\b_-^2$ is real and negative if $\lambda<0$, real and positive if $0<\lambda<1/4$ and complex with a positive real part if $\lambda>1/4$ (effective Breitenlohner-Freedman bound for this IR Lifshitz space-time).

Thus, the parameter space where \eqref{MassiveSolConstantPhi} is a stable IR fixed point (two irrelevant deformations) is $z>2$, $\lambda<0$, while it can mediate a fractionalization transition if $z>2$ and $\lambda>0$ or $1<z<2$ and $\lambda<0$.

\subsubsection{Running scalar\label{section:massiverunning}}
Finally, we turn to the case where the gauge field participates in the IR geometry at leading order and the scalar is running, with the scalar functions 
as in \eqref{EMDmassiveCouplings}. This solution exists when  $\e=\g-\d\,$:\footnote{There exists another spurious solution with $\theta=z+2$, which requires  $W_0<0$, and accordingly we discard it.}
\be \label{MassiveSolRunningScalar}
	\begin{split}
		&\ud s^2 = r^\theta\left[-\frac{\ud t^2}{r^{2z}}+\frac1{r^2}\left(L^2\ud r^2+\ud x^2+\ud y^2\right)\right],\\
		&L^2 V_0=z^2+\zeta +(\theta-2 )^2-z (\zeta +\theta-1)\,,\\
		&e^\phi=r^{\frac\theta\d}\,, \qquad \delta = \frac{\theta }{\sqrt{2 (1-z) \zeta +(\theta-2 ) \theta }},\\
		& A=\sqrt{\frac{2 (z-1)}{Z_0 (z-\zeta )}}r^{\zeta-z}\ud t\,,\qquad \gamma =\frac{\theta-2\zeta }{\sqrt{2 (1-z) \zeta +(\theta-2 ) \theta }}\\
		&\frac{W_0}{V_0Z_0}=\frac{(z-\zeta ) (2+\zeta -\theta )}{z^2+\zeta +(\theta-2 )^2-z (\zeta +\theta-1)}\,, \qquad \epsilon = \frac{-2\zeta }{\sqrt{2 (1-z) \zeta +(\theta-2 ) \theta }}\,.
	\end{split}
\ee
%Once $\gamma$, $\delta$ and the ratio $W_0/(V_0Z_0)$ have been chosen, the exponents $\theta$ can be determined uniquely from $z$, while $z$ itself is given by the root of a polynomial of degree four, so there is some degeneracy. Not all values need to statisfy the constraints 1-4 we impose below, though.
 The exponent $\zeta$ parameterizes the violation of Lifshitz scaling in the electric potential $A_t$, independently from the hyperscaling violation in the metric. In particular, this does not modify the scaling \eqref{hyperscalingviolating} of the entropy. Since $z$, $\theta$ and $\zeta$ are determined by parameters in the action, there is no free parameter left once $\gamma$, $\delta$ and the ratio $W_0/V_0Z_0$ have been chosen.

As in section \ref{section:masslessrunning}, sending $\d\to0$ ($\theta=0$) will result in a Lifshitz hyperscaling space-time, with a running scalar, where the electric potential is itself not Lifshitz-invariant, as $\zeta\neq0$. One can on the other hand take $\zeta=0$, so the electric potential is Lifshitz invariant while the metric is hyperscaling violating. Sending $z,\,\theta\to+\infty$ while keeping their ratio fixed and changing coordinates to $r^z=\rho$ is still a regular limit and makes the metric conformal to AdS$_2\times R^2$, with the gauge field invariant under the semi-local scaling. This case enjoys special dimensional reduction properties, see section \ref{app:LiftW0neq0}. Given recent interest in semi-local criticality in the symmetry-preserving phase, \cite{semilocal}, it is quite interesting that such a geometry appears also in the IR asymptotics of the symmetry-breaking phase. One can also take the limit $\zeta\to\infty$ to maintain an anomalous scaling.

\begin{figure}
\begin{tabular}{ccc}
\includegraphics[width=0.3\textwidth]{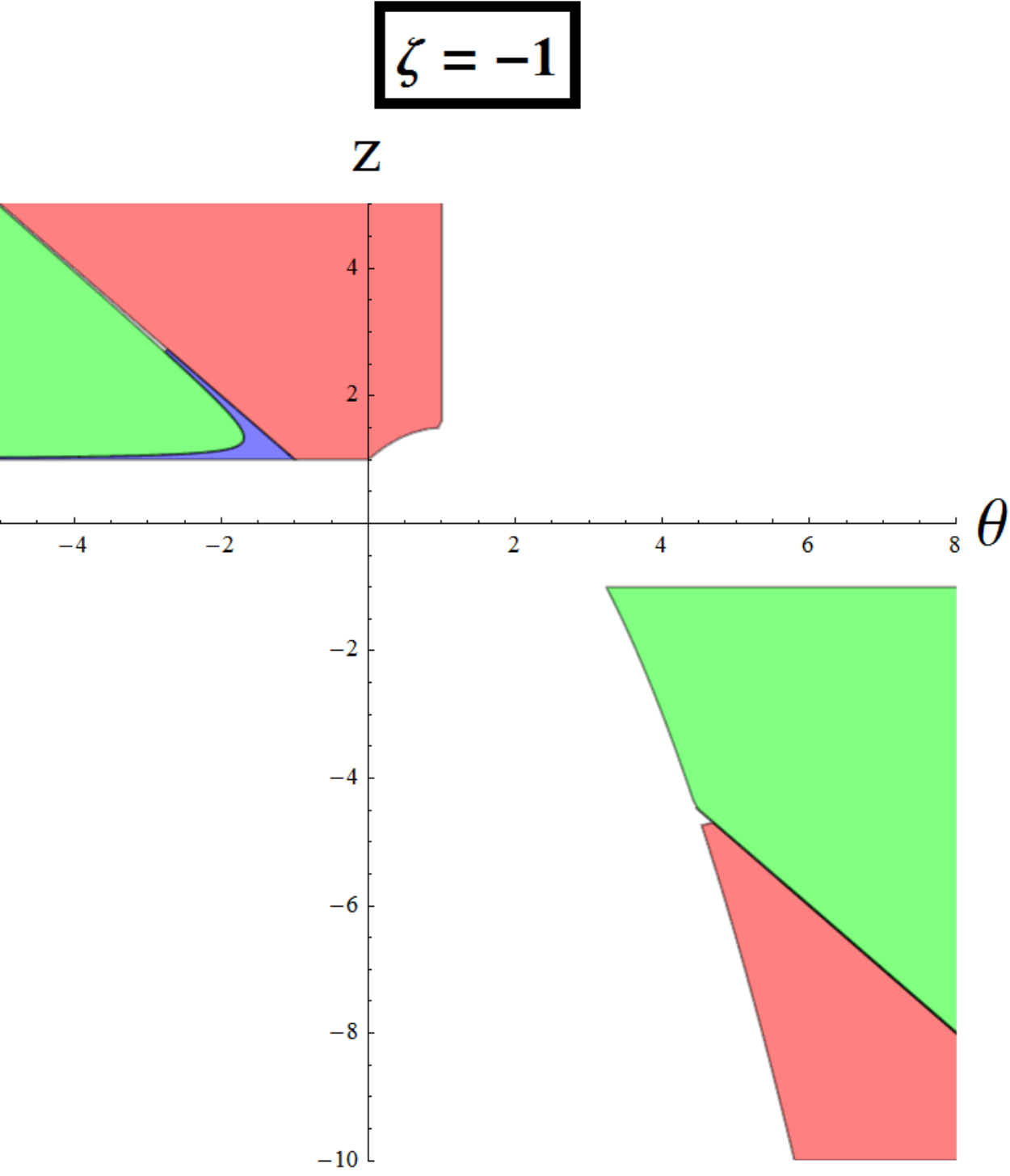}&\includegraphics[width=0.3\textwidth]{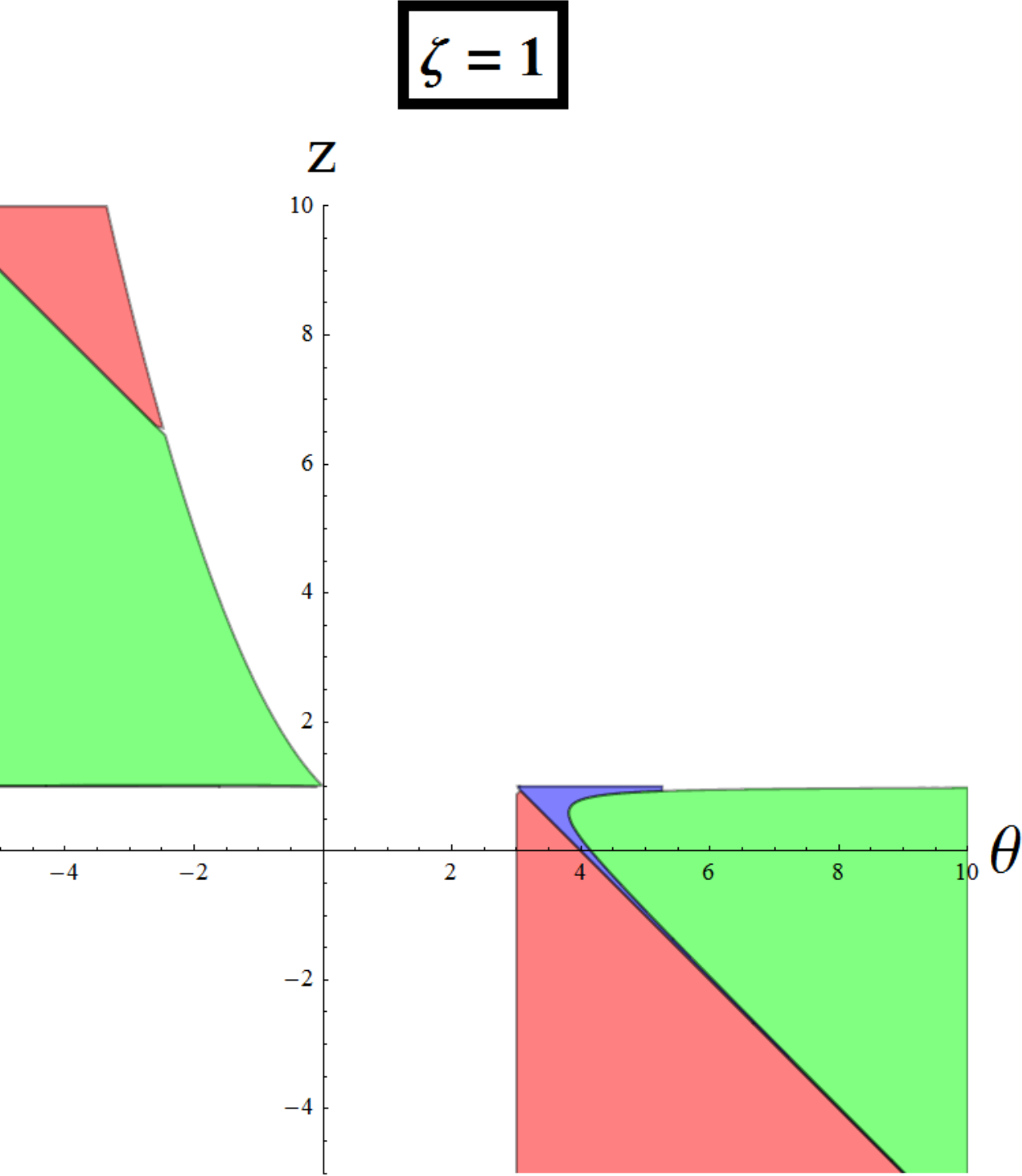}&\includegraphics[width=0.3\textwidth]{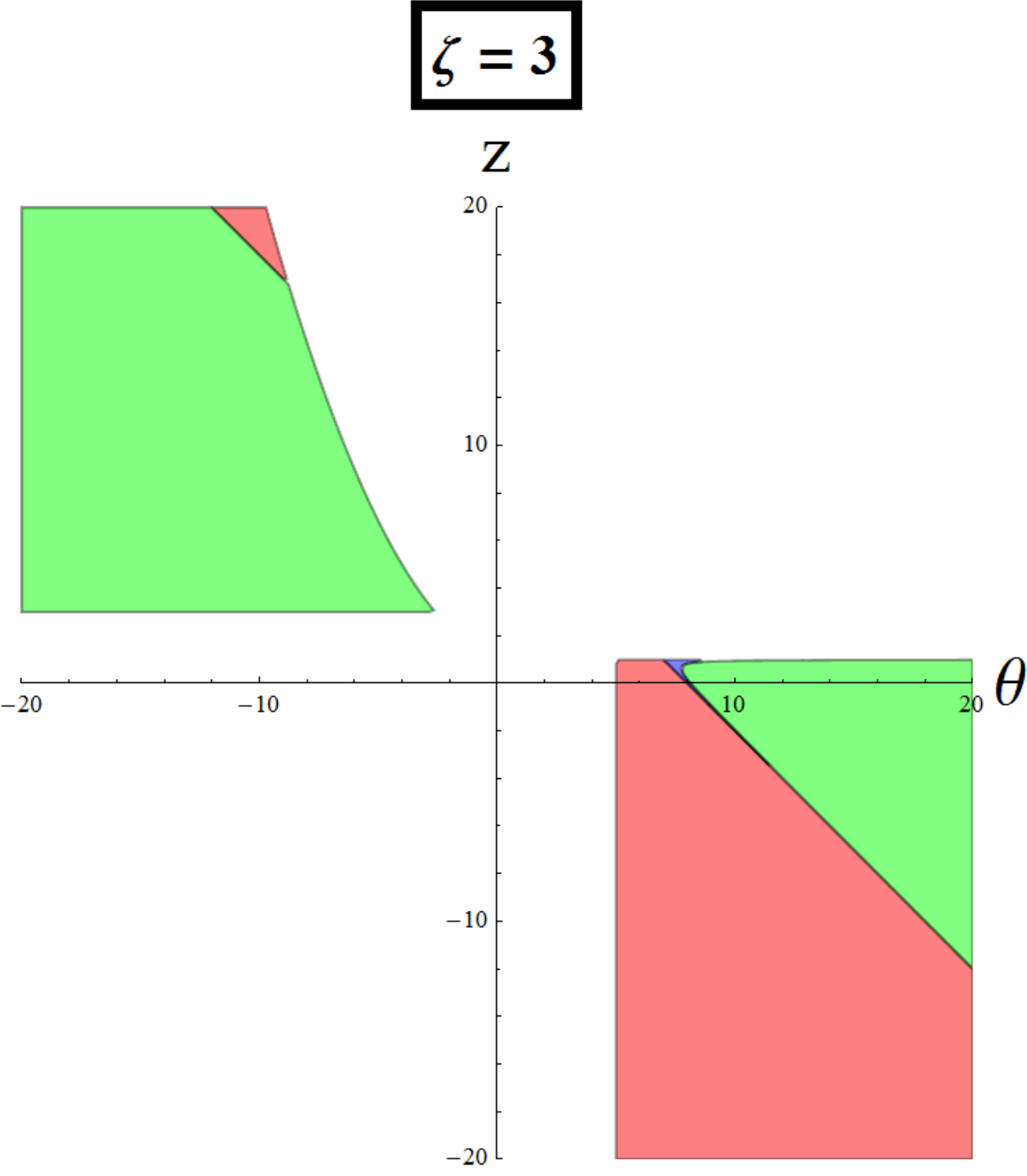}
\end{tabular}
\caption{Plots of the allowed parameter space $(\theta,z)$ for various values of the exponent $\zeta$. The upper left corner is the region where the IR is $r\to+\infty$, the lower right where it is $r\to0$. In red, we depict the region where $\beta_-$ is a real irrelevant deformation; in blue, the region where it is real and relevant; in green, the region where it is complex and relevant. In this case, the geometry \protect\eqref{MassiveSolRunningScalar} is dynamically unstable.}
\label{Fig2}
\end{figure}

We  consider the constraints one can put on the parameter space. There are five of them
\begin{enumerate}
	\item The first is to require a well-defined IR (see appendix \ref{app:hyper}), whether $r\to0$ ($2+z-\theta<0$) or $r\to+\infty$ ($2+z-\theta>0$):
\be
(\theta-2)(\theta-2z)>0\,.
\ee
	\item The second is requiring that $r$ is spacelike, $L^2>0$, which provided that $V_0>0$, implies
\be
	z^2+\zeta +(\theta-2 )^2-z (\zeta +\theta-1)>0\,.
\ee
	\item The third is that $Q^2$ is positive ($Z_0>0$), which implies
\be
	 (z-1)(z-\zeta )>0\,.
\ee
	\item The fourth is that $W_0>0$, which seems reasonable as it is related to the mass of the gauge field, and is associated to the positivity of the kinetic term of the original charged scalar,
\be
	 (z-\zeta ) (2+\zeta -\theta )<0\,.
\ee
	\item The last condition is that $\delta$ and $\gamma$ are real:
\be
2 (1-z) \zeta +(\theta-2 ) \theta>0\,.
\ee
\end{enumerate}
The allowed parameter space $(\theta,z)$ is plotted in Fig. \ref{Fig2}, for various values of $\zeta$, and is smaller than that allowed by the NEC.
We evaluate the behavior of the IR electric flux \eqref{ElFlux} as
\be
	\int_{\mathbf R^2}Z(\phi)\star F\sim r^\xi\,,\quad \xi=\theta -2-\zeta\,,
\ee
which is always vanishing in the IR. $\xi=0$ automatically implies $W_0=0$ and brings us back to the solutions of section \ref{section:masslessrunning}. In this case, one can also work out that the gauge field vanishes in the IR while $Z(\phi)$ diverges.

In \cite{Adam:2012mw}, the fact that the charged scalar did not couple directly to the gauge field kinetic term (but only the neutral scalar responsible for driving the IR asymptotics) prevented any cohesive phase unless the coupling function $Z$ stayed bounded. We can evade this constraint in our case as the charged scalar does couple \emph{via} $Z$ to the gauge field kinetic term.

 We now perturb \eqref{MassiveSolRunningScalar} linearly (details are relegated to appendix \ref{app:ChargedCohHypVio}). We are looking for irrelevant deformations. First, there is a zero mode shifting the scalar by a constant.

There is also a pair of conjugate modes $\beta_\pm$, one relevant and one irrelevant and summing to $2+z-\theta$. Their exact expressions are not illuminating and can be found in appendix \ref{app:ChargedCohHypVio}. Irrespective of where the IR is, $\b_+$ is never irrelevant. Depending on the parameters, $\beta_-$ can be irrelevant (real and negative if the IR is $r\to+\infty$, positive if it is $r\to0$) or relevant  (real and positive if the IR is $r\to+\infty$, negative if it is $r\to0$; or complex). In Fig. \ref{Fig2}, we plot these regions in the parameter space. If $\beta_-$ is complex, then it is relevant and the geometry \eqref{MassiveSolRunningScalar} is dynamically unstable.

In the red region, the amplitudes of the modes $\b=0$ and $\b_-$ can be used to tune the source of the dual scalar operator to zero in the UV and engineer spontaneous symmetry breaking.

\section{Hyperscaling violation from generalized dimensional reduction\label{app:lifts}}

In this section, we summarize how the various hyperscaling violating metrics discussed in sections \ref{section:neutralDWunbr}, \ref{section:masslessrunning} and \ref{section:massiverunning} can be obtained from  scale invariant metrics using the generalized dimensional reduction techniques of \cite{Kanitscheider:2009as,gk}.\footnote{By generalized, we mean that the number of reduced dimensions is analytically continued and exchanged against a real parameter.} Given that the extremal geometries \eqref{hyperscalingviolating} are conformal to Lifshitz, it is quite natural to expect such lifts to exist. Technical details of such uplifts can be found in \cite{gk} for the symmetry-preserving case and in appendix \ref{app:DimRed} for the symmetry-breaking case.

We shall mostly consider diagonal lifts, either along a torus or along a sphere. The Kaluza-Klein Ansatz for the metric will look like
\be
	\ud s^2_{(4+n)} = e^{-\Delta\varphi}\ud s^2_{(4)}+e^{\frac{\varphi}{\Delta}\left(1-\Delta^2\right)}\ud K^2_{(n)}\,,
\ee
where $\varphi$ coincides with the lower-dimensional scalar $\phi$ of action \eqref{1} in the absence of a higher-dimensional scalar $\Phi$, while $\Delta=\d$ in the same way. $\mathbf K_{(n)}$ is an $n$-dimensional constant curvature space, which we select to be a torus $\mathbf T^n$ or a sphere $\mathbf S^n$, in which case
\be
	\mathbf T^n:\, \Delta^2=\frac{n}{2+n}\leq1\,,\qquad\mathbf S^n:\, \Delta^2=\frac{2+n}{n}\ge1\,.
\ee
We parametrize the higher-dimensional theory as:
\be \label{KKaction}
	\tilde S=\int \ud^{4+n}\sqrt{-\tilde g}\left[\tilde R-\half\tilde\partial\Phi^2-\frac{Z_0e^{\Gamma\Phi}}{2(q+2)!}\tilde F^2_{[q+2]}-\frac{W_0e^{E\Phi}}{2(q+1)!}\tilde A^2_{[q+1]}+V_0\right],
\ee
and scan the various cases which can occur:
\begin{itemize}
	\item $Z_0=W_0=0$: the fixed point is neutral as in section  \ref{section:neutralDWunbr}.
	\item $Z_0\neq0$, $W_0=0$: the fixed point is charged with a massless gauge field, as in section \ref{section:masslessrunning}.
	\item $Z_0\neq0$, $W_0\neq0$: the fixed point is charged with a massive gauge field, as in section \ref{section:massiverunning}.
\end{itemize}

\subsection{Symmetry-preserving phases\label{app:LiftW0=0}}

We  first address the case with $Z_0=0$.
There are two different uplifts to explain the scaling of the metric \eqref{neutralDW2}, \cite{gk,Gouteraux:2011qh}:
\begin{itemize}
	\item $\delta^2<1$ ($\theta<0$), $\Phi=0$: The metric can be uplifted to a Poincar\'e-AdS in $n+4$ dimensions. Then, the higher-dimensional holographic dictionary can be reduced to define the lower-dimensional one, \cite{Kanitscheider:2009as,Gouteraux:2011qh}.
	\item $\delta^2>1$ ($\theta>0$), $\Phi=0$, $V_0=0$: The metric can be uplifted to a Ricci-flat $2$-brane with a transverse sphere $\mathbf S^n$, which can be realized either by a theory without a cosmological constant.
\end{itemize}

Turning to $Z_0\neq0$, the scaling properties of the solutions \eqref{MasslessSolEMD} with $W_0=0$ can be interpreted by lifting them to solutions of the action \eqref{KKaction}, \cite{gk}:
\begin{itemize}
	\item $\g=\d$, $\d^2<1$: $\Phi=0$, $q=0$, $V_0\neq0$. The uplifted metric is AdS$_2\times\mathbf R^{n+2}$. This has a constant scalar (so $\theta=0$) and corresponds formally to $z\to+\infty$.
	\item $\g=-\d$, $\d^2<1$: $\Phi=0$, $q=n$, $V_0\neq0$. The uplifted metric is planar AdS$_{n+2}\times\mathbf R^{2}$. Both $\theta$ and $z$ diverge, but their ratio defines a new hyperscaling violation exponent $\theta/z=-n$.
	\item $\g\d=1$, $\d^2>1$: $\Phi=0$, $q=0$, $V_0=0$. The uplifted metrics are $(n+4)$-dimensional near-extremal black branes in Einstein-Maxwell theory. In that case, one has
	\be
		\theta=\frac{2(n+2)}{(n+1)}\,,\qquad z=\frac{3-n^2}{1+n}\,,
	\ee
	and the metric is the $\Gamma=0$ limit of \eqref{NEDilMaxBrane} below.
	\item $(\g-\d)(\g+\d)>0$: $\Phi\neq0$, $q=0$, $V_0\neq0$. The uplifted metrics are $(n+4)$-dimensional Lifshitz black holes with a Maxwell charge:
	\be
		\begin{split}
		&\ud\tilde s^2=-\frac{\ud t^2}{r^{2z}}+\frac{L^2\ud r^2+\ud R^2_{(n+2)}}{r^2}\,, \quad L^2=\frac{(1+n+z) (n+2+z)}{V_0}\\
		&\theta=-n\,, \qquad z=1+\frac{2(n+2)}{\Gamma^2}\,,\\
		&e^\Phi=r^{\sqrt{2(n+2)(z-1)}}\,,\qquad \tilde A_{[1]}=\sqrt{\frac{2(z-1)}{Z_0(n+2+z)}}r^{-n-2-z}\ud t\,.
		\end{split}
	\ee
	\item  $(\g-\d)(\g+\d)>0$: $\Phi\neq0$, $q=n$, $V_0\neq0$.  The uplifted metrics are Lifshitz black $n$-branes with an electric $[n]$-charge:
	\be
		\begin{split}
		&\ud\tilde s^2=\frac{-\ud t^2+\ud R^2_{(n)}}{r^{2z}}+\frac{L^2\ud r^2+\ud R^2_{(2)}}{r^2}\,, \quad L^2=\frac{(1+(n+1)z) (2+(n+1)z)}{V_0}\\
		&\theta=-nz\,, \qquad z=1+\frac{4}{\Gamma^2}\,,\\
		&e^\Phi=r^{2\sqrt{z-1}}\,,\qquad \tilde A_{[n+1]}=\sqrt{\frac{2(z-1)}{Z_0(2+(n+1)z)}}r^{-2-(n+1)z}\ud t\wedge\ud R_{(n)}\,.
		\end{split}
	\ee
	\item $(\g-\d)(\g+\d)<0$: $\Phi\neq0$, $q=0$, $V_0=0$. The uplifted metrics are near-extremal dilatonic black $2$-brane solutions with an electric Maxwell charge:
	\be \label{NEDilMaxBrane}
		\begin{split}
		&\ud\tilde s^2=-r^{\frac{4(n-1)(n+1)}{2(n+1)+(n+2)\Gamma^2}}\ud t^2+r^{\frac{-4(n-1)}{2(n+1)+(n+2)\Gamma^2}}\left(L^2\ud r^2+\ud R_{(2)}^2+r^2\ud\Omega^2_{(n)}\right)\,, \\
		&\theta=n+1+z\,, \quad z=\frac{2(3-n^2)+(n+2)\Gamma^2}{2(n+1)+(n+2)\Gamma^2}\,,\quad L^2=\frac{n(n-1)}{V_0}\,,\\
		&e^\Phi=r^{\frac{2(n-1)(n+2)\Gamma}{2(n+1)+(n+2)\Gamma^2}}\,,\qquad \tilde A_{[1]}=\sqrt{\frac{2(n+2)}{Z_0\left[2(n+1)+(n+2)\Gamma^2\right]}}r^{n-1}\ud t\,.
		\end{split}
	\ee
	Note that for $n=0$ is ill-defined, as it imposes $V_0=0$ and leaves $L^2$ undetermined.
	\item
	 $(\g-\d)(\g+\d)<0$: $\Phi\neq0$, $q=n$, $V_0=0$.:
	  \be
		\begin{split}
		&\ud\tilde s^2=-r^{\frac{-4}{2(n+1)+(n+2)\Gamma^2}}\ud t^2+r^{\frac{4(n+1)}{2(n+1)+(n+2)\Gamma^2}}\left(L^2\ud r^2+\ud R_{(2)}^2+r^2\ud\Omega^2_{(n)}\right)\,, \\
		&\theta=1+z(n+1)\,, \qquad z=\frac{2(3+2n)+(n+2)\Gamma^2}{2(n+1)+(n+2)\Gamma^2}\,,\\
		&e^\Phi=r^{\frac{2(n+2)\Gamma}{2(n+1)+(n+2)\Gamma^2}}\,,\qquad L^2=\frac{zn(zn-1)}{V_0}\,,\\
		&\tilde A_{[n+1]}=\sqrt{\frac{4(n+2)}{Z_0\left[2(2n^2+2n-2)+(n^2+n-2)\Gamma^2\right]}}r^{nz-1}\ud t\wedge\ud\Omega_{(n)}\,.
		\end{split}
	\ee
Note that the electric flux is wrapped around the transverse sphere, but can be dualized to a magnetic one wrapping the $2$-brane.
\end{itemize}
A last, quite interesting possibility is the case where the solution descends from a (neutral) AdS$_5$ solution by a circle reduction. Then $\g=\pm\sqrt{3}$ and $\d=\pm1/\sqrt{3}$. The reduction is not generalized, but the values taken by $\g$ and $\d$ imply that $z=3$ and $\theta=1$, \cite{Charmousis:2012dw}.\footnote{For more details one how to set up the holographic dictionary, see \cite{Gouteraux:2011qh}.}  In four dimensions, this is precisely the value of $\theta$ where hyperscaling violating metrics exhibit logarithmic violation of the area law of the entanglement entropy, \cite{Ogawa:2011bz,sachdev,dong}. One might say that the dual theory is one-dimensional, though this might be a little misleading as the gauge carriers do not scale as in one spatial dimension. However, the value $z=3/2$ appears in certain gauge theories describing non-Fermi liquids (together with $\theta=1$), \cite{sachdev} , not $z=3$.

One may hope that combining a non-diagonal reduction along a circle and a diagonal reduction along a torus might allow more freedom in the dynamical exponent. This setup was studied in \cite{Gouteraux:2011qh}, and yields $\theta=1-n/2$, $z=3+n/2$ for an $n$-dimensional torus. Requiring $\theta=1$ implies $n=0$ and brings us back to the circle reduction.

\subsection{Symmetry-breaking phases\label{app:LiftW0neq0}}

As for $W_0=0$ (see the previous subsection), one may still interpret the scaling properties of the solutions \eqref{MassiveSolRunningScalar} with $W_0\neq0$ using dimensional reduction tools. It does not look like a non-diagonal reduction along an $\mathbf S^1$ from an AdS theory can give a massive gauge field with $\theta=1$ in the reduced metric in our setup (though see \cite{LifshitzFromAdS}). Turning to diagonal reductions, one finds that the scaling properties of the solution can be reinterpreted as scale invariance in the higher-dimensional setup \eqref{KKaction}:
\begin{itemize}
	\item $\g=\d$, $\e=0$: $\Phi=0$, $q=0$, $V_0\neq0$. The uplifted metric is simply the massive Lifshitz solution in $n+4$ dimensions, \cite{taylor}. In the lower-dimensional setup, one finds that
	\be
		\begin{split}
	&\ud \tilde s^2=-{\ud t^2\over r^{2z}}+{L^2\ud r^2+\ud R^2_{(n+2)}\over r^2},\\
	&\tilde A_{[1]}=\sqrt{\frac{2(z-1)}{Z_0z}}r^{-z}\ud t\,, \qquad W_0=\frac{z(n-2)}{L^2},  \\
	&\theta=-n\,,\qquad V_0L^2=z^2+(n+1)z+(n+2)^2.
		\end{split}
	\ee
	\item $\g=-\d$, $\e=-2\d$: $\Phi=0$, $q=n$, $V_0\neq0$. The uplifted metric is a massive Lifshitz $n$-brane, with an electric flux wrapped around the brane, \cite{taylor}.
	\be
		\begin{split}
	&\tilde \ud s^2={L^2\ud r^2 +\ud R_{(2)}^2\over r^2}+{-\ud t^2+\ud R_{(n)}^2\over r^{2z}}\,,\\
	&\tilde A_{[n+1]}=\sqrt{\frac{2(z-1)}{(n+1)z}}r^{-(n+1)z}\ud t\wedge\ud R_{(n)}\,,\quad W_0=\frac{2(n+1)z}{L^2}\,,\\
	&\theta=-nz\,,\quad V_0L^2=(n+1)^2z^2+(3n+1)z+4\,.
		\end{split}
	\ee
	\item $\Phi\neq0$, $q=0$, $V_0\neq0$: the metric uplifts to a Lifshitz solution with a running scalar and a massive gauge field:
	\be
		\begin{split}
		&\ud\tilde s^2=-\frac{\ud t^2}{r^{2z}}+\frac{\ud R^2_{(n+2)}+ L^2\ud r^2}{r^2}\,,\quad e^\Phi=r^{\Gamma(z-1)}\,,\\
		&\theta=-n\,,\quad L^2=\frac{Z_0}{4W_0}\left[2(n+2)+\Gamma^2(z-1)\right]\left[2z+\Gamma^2(z-1)\right],\\
		&\tilde A_{[1]}=\sqrt{\frac{4(z-1)}{Z_0\left[2z+(z-1)\Gamma^2\right]}}\rho^{-z+\frac{\Gamma}{2}(z-1)}\,.
		\end{split}
	\ee
	\item $\Phi\neq0$, $q=n$, $V_0\neq0$: the metric uplifts to a Lifshitz $n-$brane with a running scalar and a massive $n$-form:
	\be
		\begin{split}
		&\ud\tilde s^2=-\frac{\ud t^2+\ud R^2_{(n)}}{r^{2z}}+\frac{\ud R^2_{(2)}+ L^2\ud r^2}{r^2}\,, \quad e^\Phi=r^{-\Gamma(z-1)}\,,\\
		&\theta=-nz\,,\quad L^2=\frac1{2V_0}\left[8+2(3n+1)z+2(n+1)^2z^2+\Gamma^2(z-1)^2\right],\\
		&\tilde A_{[n+1]}=\sqrt{\frac{4(z-1)}{Z_0\left[2(n+1)z+(z-1)\Gamma^2\right]}}r^{-(n+1)z-\frac{\Gamma^2}{2}(z-1)}\ud t\wedge\ud R_{(n)}\,,\\
		 &\frac{W_0}{Z_0V_0}=\frac{\left[4+(1-z)\Gamma^2\right]\left[2(n+1)z-(1-z)\Gamma^2\right]}{2\sqrt{8+2(3n+1)z+2(n+1)^2z^2+\Gamma^2(z-1)^2}}\,.
		\end{split}
	\ee
	\item $\Phi\neq0$, $q=0$, $V_0=0$: the metric uplifts to a hyperscaling violating $2-$brane with a running scalar and a massive electric gauge field:
	\be
		\begin{split}
		&\ud\tilde s^2=r^{\frac{2\theta}{n+2}}\left[-\frac{\ud t^2}{r^{2z}}+\frac{\ud R^2_{(2)}+L^2\ud r^2}{r^2}+\ud\Omega^2_{(n)}\right]\,, \quad L^2=\frac{n(\theta-2-z)}{V_0},\\
		&e^\Phi=r^{a}\,,\quad a^2=-4 (2+n)-2 (1+n) z-2 z^2+(2 (3+n)+2 z) \theta -\frac{2 (1+n) \theta ^2}{2+n},\\
		&\tilde A_{[1]}=\sqrt{\frac{(z-1)^2}{Z_0(2+n-\theta)(\theta-2-z)}}r^{\frac{(2+n-\theta)(2+z-\theta)}{(z-1)}}\ud t\,,\\
		&\frac{W_0}{Z_0V_0}=\frac{(2+n-\theta)(2+z-\theta)(1+n+z-\theta)}{n(z-1)^2}\,, \quad \Gamma=E+\sqrt{\frac{2}{n+2}}\frac\theta{a}\,,\\
		&E^2=\frac{2 (2+n) \left(z+z^2+n (2+z-\theta )+(-2+\theta )^2-z \theta \right)^2}{(-1+z)^2 a^2}\,.
		\end{split}
	\ee
		\item $\Phi\neq0$, $q=n$, $V_0=0$: the metric uplifts to a Lifshitz $n-$brane with a running scalar and a massive $n$-form:
	\be
		\begin{split}
		&\ud\tilde s^2=r^{\frac{2\theta}{n+2}}\left[-\frac{\ud t^2}{r^{2z}}+\frac{\ud R^2_{(2)}+ L^2\ud r^2}{r^2}+\ud\Omega^2_{(n)}\right]\,, \quad L^2=\frac{nz(\theta-2-z)}{V_0}\,,\\
		&e^\Phi=r^{a},\quad a^2=-8-2 (1+2 n) z-2 (1+n) z^2+(6+2 (1+n) z) \theta -\frac{2 (1+n) \theta ^2}{2+n}\\
		&\tilde A_{[n+1]}=\sqrt{-\frac{(-1+z)^2}{Z_0(2+z-\theta ) (2+n z-\theta )}}r^{\frac{(2+z-\theta ) (2+n z-\theta )}{-1+z}}\ud t\wedge\ud \Omega_{(n)}\,,\\
		&\frac{W_0}{Z_0V_0}=\frac{(2+z-\theta ) (2+n z-\theta ) (1+z+n z-\theta )}{n (-1+z)^2 z}\,.
		\end{split}
	\ee
	Note that the electric flux is wrapped around the transverse sphere, but there is no electromagnetic-duality to dualize is to a magnetic flux wrapped around the $2$-brane: this symmetry of the equations of motion is broken by the mass term of the gauge field.
\end{itemize}

\section{Outlook}

In this work, we have given a unifying view of the classification of the IR asymptotics of holographic theories using the concept of Effective Holographic Theories, as advocated first in \cite{cgkkm,gk}. We have considered standard Abelian scaling symmetry but have enlarged our scope by allowing broken U(1) symmetries.
We have found all IR scaling, homogeneous and translation-invariant asymptotics and studied their perturbations. We have classified them in terms of whether they describe  cohesive or fractionalized phases, whether the symmetry-breaking effects are leading or subleading in the IR and whether the scale invariant fixed points can mediate a fractionalization transition via a bifurcation in the RG flow.

There is a non-trivial interplay between hyperscaling preserving and hyperscaling violating solutions. Hyperscaling preserving solutions appear as fixed points
and have AdS$_4$, AdS$_2\times R^2$ or Lifshitz geometries. The hyperscaling violating solutions appear as fixed lines and contain an explicit scale (responsible for hyperscaling violation). The fixed line can be interpreted as changing that scale, or equivalently changing the charge density (or chemical potential).
This structure, albeit more complex is expected to appear in more general EHTs, containing several scalars and U(1) symmetries.

Our surprising result is that scaling (quantum critical) behavior is generic in holographic theories and that quantum critical theories appear generically as fixed lines rather than fixed points. A possible connection of this fact to recent data on cuprate superconductors has been addressed in \cite{kkp}.

There are several directions that open up:

\begin{itemize}
	\item We have introduced a new scaling exponent $\zeta$ in section \ref{section:massiverunning}, which parameterizes the violation of the Lifshitz scaling in the electric potential, independently from the violation of hyperscaling in the metric. It would be very interesting to find a connection between this exponent in the bulk and scaling in the dual field theory.
	\item Understanding the structure of flows between the critical solutions we find and map, is of importance.
This will put an order in this collection of scaling geometries. Some specific cases have been studied in \cite{Hartnoll:2011pp,Hartnoll:2012pp,Adam:2012mw}, but as we have described here, this is just the tip of an iceberg.
\item It is interesting to investigate new effects that may appear after the inclusion of more order parameters/symmetries. In particular new effects are expected when what acts as parameters in single scalar models, may turn into varying parameters in multi-scalar models. Indeed that is what happens in tachyon models for chiral symmetry breaking of QCD, \cite{chi}, where new phenomena like conformal transitions can appear, \cite{jk}.
\item  The special class of hyperscaling violating metrics conformal to AdS$_2\times R^2$ have been singled out in \cite{gk}, because they have vanishing entropy at extremality (as do all hyperscaling violating metrics) and display a ``semi-local'' quantum criticality (time scales, space does not, but Green's functions are still momentum-dependent), \cite{semilocal}. They have received a lot of attention recently, and were shown to have ``fermionic'' properties. The previous studies have focussed on the symmetry-preserving phase, but this class of geometries also appears in the symmetry-breaking phases. It would be interesting to investigate further their properties.

\item Logarithmic violation of the area law of the entanglement entropy has been tied to the presence of Fermi surfaces in \cite{Ogawa:2011bz,sachdev,dong}. This does not depend on the theory considered, but rather on the metric itself, and more precisely on the value of $\theta$ (and not $z$). In particular, the geometries presented here, for instance \eqref{MasslessSolEMDPower} or \eqref{MassiveSolRunningScalar}, can accommodate this value. However, they are supposed to model superfluid phases, where the fermionic degrees of freedom have condensed. How is it then that a Fermi surface should still be present? For fractionalized phases, the fractionalized fermion would still display a Fermi surface,\footnote{We would like to thank A. Balatsky for discussions on this specific point.} but not for the cohesive phase of \eqref{MassiveSolRunningScalar}. This leads one to wonder whether weakly-coupled concepts such as the Fermi surface are still applicable in this strongly-coupled context, and even make sense.

\item Given the ubiquitousness of hyperscaling violating metrics in Effective Holographic models, it is important to investigate their stability under quantum or stringy corrections. The former case has already been studied in a number of works, \cite{stability} and it was found that magnetic versions of these metrics are unstable to quantum effects and develop again an AdS$_2\times R^2$ throat. The latter case is still largely unexplored. Lifshitz metrics have been showed to be stable, \cite{Adams:2008zk}, while inclusion of certain quadratic operators maintaining second-order field equations preserved the two-derivative scaling symmetries, \cite{Charmousis:2012dw}.
\end{itemize}

\vskip 0.5cm

 \addcontentsline{toc}{section}{Acknowledgements}
\section*{Acknowledgements}

We would like to thank A. Balatsky, S. Hartnoll, J. Hartong, L. Huijse and B. Withers for interesting discussions, as well as the organizers of the program ``The holographic way: string theory, gauge theory and black holes'' held at NORDITA for a stimulating environment during the final stages of this work.

This work was in part supported by grants
 PERG07-GA-2010-268246, PIEF-GA-2011-300984, the EU program ``Thales'' ESF/NSRF 2007-2013, and by the European Science
Foundation ``Holograv" (Holographic methods for strongly coupled systems) network.
It has also been co-financed by the European Union (European Social Fund, ESF) and
Greek national funds through the
 Operational Program ``Education and Lifelong Learning'' of the National Strategic
 Reference Framework (NSRF) under
 ``Funding of proposals that have received a positive evaluation in the 3rd and 4th Call of ERC Grant Schemes''.

\vskip 0.5cm

\section*{Note Added}

While this work was being submitted we received reference \cite{kasa} which has some overlap with section \ref{section:massiverunning} our work. Shortly after our submission, \cite{nbi} appeared with similarities also to section \ref{section:massiverunning}.

\newpage
 \addcontentsline{toc}{section}{Appendices}
  \renewcommand{\theequation}{\thesection.\arabic{equation}}

\appendix

\settocdepth{section}

\section{Ansatz and field equations\label{app:Ansatz}}

We consider a metric in a general frame
\be
ds^2= {B(r)\ud r^2}-D(r)\ud t^2+C(r)\left(\ud x^2+dy^2\right),
\label{8}\ee
and with non-trivial gauge field, $A_t(r)$ and scalar $\phi(r)$.

Using primes to denote radial derivatives, the scalar curvature of this metric is
\be \label{RicciScalarGenMetric}
R={1\over B}\left[{1\over 2}\left({C'\over C}-{D'\over D}\right)^2+{B'\over B}\left({C'\over C}+{D'\over 2D}\right)-2{C''\over C}-{D''\over D}\right].
\ee
Specifying to an electric Ansatz for the gauge field, the equations of motion read
\be
\left(Z\sqrt{C^2\over DB}A_t'\right)'=W\sqrt{BC^2\over D}A_t
\label{MaxEqmassive}\ee
\be
{D'\over D}{C'\over C}+{1\over 2}{C'^2\over C^2}+{Z\over 2}{A_t'^2\over D}
-{BW\over 2}{A_t^2\over D}-BV-{1\over 2}\phi'^2=0
\label{Eeq1massive}\ee
\be
{C''\over C}+{1\over 2}\phi'^2-{1\over 2}\left({B'\over B}+{D'\over D}\right){C'\over C}
-{1\over 2}{C'^2\over C^2}+{BWA_t^2\over 2D}=0
\label{Eeq2massive}\ee
\be
2{D''\over D}-2{C''\over C}-{D'\over D}\left({D'\over D}+{B'\over B}-{C'\over C}\right)
+{B'\over B}{C'\over C}-{2ZA_t'^2\over D}
-{2BWA_t^2\over D}=0
\label{Eeq3massive}\ee
\be
\sqrt{1\over DBC^2}\partial_r\left(\sqrt{DC^2\over B}\phi'\right)+V'+{Z'\over 2DB}A_t'^2+{W'\over 2D}A_t^2=0
\label{DilEqMassive}\ee
Only four of them are independent, as the scalar equation of motion \eqref{DilEqMassive} can be obtained from the others.

From these equations, one can obtain the following conserved charge
\be\label{extrCharge}
\mathcal Q=\frac{C}{\sqrt{BD}}\left[ZAA'-C\left(\frac{D}C\right)'\right].
\ee
This charge connects horizon to boundary data, and when it evaluates to zero, signals extremality, \cite{GubserNellore,cgkkm,Hartnoll:2011pp}. Note that interestingly, the precise matter content (massive gauge field, charged complex scalar, electron star) does not affect its definition.

To connect with hyperscaling (violating) solutions, we can look for solutions of the form:
\be
	B(r)=L^2 r^{\theta-2}, \qquad D(r)=r^{\theta-2z},\qquad C(r)=r^{\theta-2},\qquad A(r)=Q r^{\zeta-z}\,.
	\label{HyperscalingAnsatz}
\ee
All scaling exponents $z$, $\theta$ and $\zeta$ can be determined in terms of the scalar exponents, $\ga$, $\da$, $\epsilon$ and the action parameters $Z_\star$, $W_\star$ and $V_\star$ or $Z_0$, $W_0$ and $V_0$, which is reminescent \cite{KT} of the attractor mechanism at work for solutions with constant scalars.

\section{Some properties of hyperscaling violating metrics \label{app:hyper}}

In this appendix, we collect some properties of hyperscaling violating metrics.
 In arbitrary $d+2$ dimensions, hyperscaling violating metrics can be written
	\be\label{UVconstraint}
		\ud s^2_{(d+2)}=r^{\frac{2\theta}d}\left[-\frac{\ud t^2}{r^{2z}}+\frac{L^2\ud r^2+\ud R^2_{(d)}}{r^2}\right].
	\ee
To have a well-defined IR, we should require
\be \label{NoIRBoundary}
	(\theta-d)(\theta-dz)>0\,.
\ee
The IR can be located either at $r\to0$ or $r\to+\infty$,\footnote{One should however keep in mind that the $r$ coordinate in \eqref{UVconstraint} will generically not coincide with the radial coordinate describing the full RG flow from AdS in the UV. The space-time \eqref{UVconstraint} will typically be obtained by a scaling limit, which implies changing radial coordinates. In the full picture, the IR may well sit at a finite value of the radial coordinate.} and this is determined by where the $(x,t)$ metric elements vanish:
\be \label{IRconstraint}
\begin{split}
	r\underset{IR}{\to}+\infty&\quad \theta<d\,,\quad \theta<dz\,,\\
r\underset{IR}{\to}0&\quad \theta>d\,,\quad \theta>dz\,.
\end{split}
\ee
Independently of a precise embedding of \eqref{UVconstraint}, and anticipating on appendix \ref{app:LinPert}, it can be seen by calculating the extremality charge \eqref{extrCharge} that perturbations with the Ansatz \eqref{perturbationsconstant} or \eqref{perturbations} and a mode equal to $d+z-\theta$ will generically correspond to flows to finite temperature. This is also supported by the fact that whenever an exact analytical completion of  \eqref{UVconstraint} to finite temperature exists, it reduces to that particular mode for small temperatures. That perturbation also helps defining the UV, since it should vanish there. As a consequence, let us add the following constraint to \eqref{IRconstraint}:
\be \label{ConsistentT}
\begin{split}
	r\underset{IR}{\to}+\infty&\quad d+z-\theta>0\,,\\
r\underset{IR}{\to}0&\quad d+z-\theta<0\,.
\end{split}
\ee
In turns out that the constraints \eqref{IRconstraint} as well as other ``natural'' constraints (such as $L^2>0$ or $Q^2>0$ if hyperscaling violation is supported by extra matter fields, see sections \ref{section:masslessrunning}, \ref{section:neutralDW}, \ref{section:massiveSolRunningScalar}) will all yield \eqref{ConsistentT}. Nonetheless, there is no proof that it should always be the case. On top of that, one may wish to impose the Null Energy Condition (NEC), which reads for the two null vectors $(N^t=L e^{z-\frac\theta d}\,,\, N^r=\frac1L  r^{1-\frac\theta d}\cos\psi\,,\,N^x=\frac1L  r^{1-\frac\theta d}\sin\psi\,)$ where $\psi=(0,\pi/2)$, \cite{dong}:
	\be \label{NEC}
		\begin{split}
			&(d-\theta)(d(z-1)-\theta)\geq0\,,\\
			&(z-1)(d+z-\theta)\geq0\,.
		\end{split}
	\ee
Combining these two sets of constraints, \eqref{UVconstraint} and \eqref{NEC} yields:
\be \label{NEC+UV}
	\theta\leq0\, \&\&\, z\geq1\quad ||\quad 0<\theta<d\,\&\&\,z\geq1+\frac\theta{d}\,.
\ee
Note that in any case, $z\geq1$.
	We will also be interested to know how the NEC  translates after uplifting with the Kaluza-Klein Ansatz:
	\be
		\ud \tilde s^2 = e^{-\da\phi}\ud s^2_{(4)}+e^{\frac{\phi}{\da}(1-\da^2)}\ud \Omega^2_{(n)}, \quad \da^2=\frac{n}{n+2},
	\ee
	where $n$ is the number of compact dimensions, $\phi$ the Kaluza-Klein scalar and for simplicity $\Omega_{(n)}$ is an $n$-sphere. Then, projecting the Einstein tensor on the null vector, the higher-dimensional NEC reads in terms of the lower one:
	\be
		\tilde G_{\mu\nu}N^\mu N^\nu= \left(G_{\mu\nu}-\frac12\partial_\mu\phi\partial_\nu\phi\right)N^\mu N^\nu\,,
	\ee
	where the indices $\mu,\nu=0\ldots3$ span the non-compact dimensions.
	
The sign of $\theta$ determines whether the geometry has a curvature singularity in the IR:
\be
	R\sim r^{-\frac2d\theta}, \qquad R_{\lambda\mu\nu\rho}R^{\lambda\mu\nu\rho}\sim r^{-\frac4d\theta}
\ee
where $R$ is the Ricci scalar calculated from \eqref{RicciScalarGenMetric}. The curvature singularity sits either at $r=0$ ($\theta>0$) or at $r=+\infty$ ($\theta<0$). In previous work, \cite{cgkkm}, we argued that one should have the curvature singularity in the IR, so as not to spoil the interpolation to a UV solution with appropriate (AdS) asymptotics. However, a numerical solution was presented recently with $0<\theta<2$, \cite{Ogawa:2011bz}, which interpolates to AdS UV boundary conditions before the curvature invariants can diverge.

Even if the curvature invariants are regular there, the IR space-time may still suffer from diverging tidal forces. Then, strings propagating in this background may become infinitely excited, preventing any resolution of this null singularity by including the full spectrum of string states, \cite{horway}. It turns out that the only value of $z$ compatible with \eqref{NEC+UV} where these null singularities do not occur is $z=1+\frac\theta{d}$, \cite{shagh}, which saturates one of NEC inequalities.

There has been a lot of recent activity around a specific value for $\theta$, which lies in the allowed range \eqref{NEC+UV}: when $\theta=d-1$, the system is effectively $(d-1)$-dimensional and the area law for the entanglement entropy is logarithmically violated. This has been argued to signal a ``hidden'' Fermi surface where the charge carriers are gauge-variant operators associated to horizon degrees of freedom, \cite{Ogawa:2011bz,sachdev,dong}. In $d=2$ dimensions, one is then led to select the values $\theta=1$, $z=\frac32$.

Can these values be realized in the geometries examined in the main text? We  consider in turn the massless geometries of section \ref{section:masslessrunning} and the massive geometries of section \ref{section:massiverunning}. In the former, setting $\theta=1$ yields $\gamma=3\delta$, and $z$ still depends on $\delta$, \eqref{MasslessSolEMD}, \cite{sachdev}:
\be
	\g=3\d \Rightarrow \theta=1,\quad z=\frac{3\delta^2+1}{2\delta^2}\,.
\ee
Varying $\delta$, z takes values in the real line and is bounded from below, $3/2<z$, with the lower-bound saturated when both $\g$ and $\d$ diverge.\footnote{We would like to thank Jelle Hartong for bringing this point to our attention.} Turning to the latter, \eqref{MassiveSolRunningScalar}, when $\theta=1$,
\be \label{Theta=1}
	z=\frac{1+\gamma  \delta }{\delta  (\gamma -\delta )}\,,\quad W_0=-\frac{V_0 Z_0(-\gamma +\delta ) (-\gamma +3 \delta ) \left(2+\gamma ^2+\delta ^2\right)}{2 \left(2+\gamma ^2+2 \gamma  \delta +\delta ^2+5 \gamma ^2 \delta ^2-6 \gamma  \delta ^3+3 \delta ^4\right)}
\ee
One recovers the massless case by setting $\gamma=3\delta$. Setting further $z=3/2$ implies a negative mass: $W_0<0$.\footnote{This result was pointed out to us by Jelle Hartong.} As $W_0$ should be thought of as the charge squared of our original complex scalar, this leads to some inconsistency. To conclude, it does not appear possible to satisfy both constraints coming from logarithmic violation of the area law of the entanglement entropy and IR regularity in the minimal setup \eqref{1}.

\section{Classification of QC asymptotics in the broken symmetry case. \label{conv}}

We write the metric as:
\be
ds^2= {L^2B(r)\ud r^2}-D(r)\ud t^2+C(r)\left(\ud x^2+dy^2\right).
\label{oo}\ee
We take again the asymptotic IR forms
\be
V=V_0 e^{-\d\phi}\sp Z=Z_0 e^{\g\phi}\sp W=W_0 e^{\e\phi}
\label{1o}\ee
but with general $\d,\,\g,\,\e$,
and make the Ansatz (if $\t<2$, $\t<2z$ then the UV is at $r=0$)
\be\label{2o}
\begin{split}
	&D(r)=r^{\theta-2z}\left(1+\sum_{n=1}\mathfrak d_nr^{n\a}\right),\quad B(r)=r^{\theta-2}\left(1+\sum_{n=1}\mathfrak b_nr^{n\a}\right),\\
	&C=r^{\t-2},\quad e^\phi = r^{a_0}\left(1+\sum_{n=1}\varphi_nr^{n\a}\right),\quad A_t=Qr^{c_0}\left(1+\sum_{n=1}\mathfrak a_nr^{n\a}\right).
\end{split}
\ee
The field equations should determine uniquely both the leading order as well as the subleading power series.

The equations (\ref{MaxEqmassive})-(\ref{Eeq3massive}) imply
\be
{ZA_t'^2\over D}={k_1\over r^2}+\cdots\sp {BWA_t^2\over D}={k_2\over r^2}+\cdots\sp BV={k_3\over r^2}+\cdots
\label{4o}\ee
with
\be
k_1-k_2-2k_3=a_0^2+(\t-2)(4z-3\t+2)
\label{5o}\ee
\be
k_2=-a_0^2+(\t-2)(\t+2-2z)
\label{6o}\ee
\be
2(k_1+k_2)=4(z^2-z\t+z+\t-2)
\label{7o}\ee
to leading order.
These can be solved for $k_i$ as
\be
k_1=a_0^2+2z(z-1)-\t(\t-2)
\label{8o}\ee
\be
k_2=-a_0^2+(2+\t-2z)(\t-2)
\label{9o}\ee
\be
2k_3=a_0^2+2(z^2+z+4)-2\t(z+3)+\t^2
\label{10o}\ee
The gauge field equation reads
\be
c_0 (-2 + c_0 + a_0 \g + z) {Z_0\over L} r^{-3 + c_0 + a_0 \g + z} =L W_0  r^{-3 + c_0 + a_0 \e + \t + z}.
\label{12o}\ee
If the potential term is leading  ($k_3\not=0,BV\sim 1/r^2$) then
\be
a_0\delta=\t \sp k_3=L^2V_0e^{-\d\phi_0}\,.
\label{11o}\ee
There are two possibilities here:

\begin{enumerate}

\item Both terms in (\ref{12o}) are leading. In that case
\be
\theta=a_0(\g-\e)\sp L^2=c_0(c_0+a_0\g+z-2){Z_0\over W_0}\,.
\label{14o}\ee

Note also that there is a compatibility condition for the first two equations in (\ref{4o}).
Its implications depend on the case. Moreover, combining \eqref{11o} with \eqref{14o} necessarily requires setting $\epsilon=\gamma-\delta$ if the dilaton is to be non-trivial ($a\neq0$).

\item Only the first term in (\ref{12o}) is leading implying
\be
c_0(c_0+a_0\g+z-2)=0\sp (\t-2)(\t+a_0(\e-\g))>0
\label{1oo}\ee
where the inequality is necessary so that the mass term is subleading in the deep IR.

\end{enumerate}

We are then left with algebraic equations.
We may now consider what happens  depending on whether $k_i$ are zero or not.
The relevant equations are (\ref{4o}), (\ref{8o}-\ref{10o}) and either (\ref{14o}) or (\ref{1oo}).

\subsection{$\prod_{i=1}^3 k_i\not=0$}

From the equations (\ref{4o}) we have
\be
c_0^2Q^2Z_0=k_1\sp L^2Q^2W_0=k_2\sp L^2V_0=k_3
\label{76o}\ee
\be
\t=a_0\d\sp c_0=-z-{a_0\e\over 2}\sp \t=a_0(\g-\e)
\label{77o}\ee
that imply
\be
a_0(\e-\g+\d)=0\, \Rightarrow \, \epsilon=\g-\d
\label{15o}\ee
for a running scalar.
We can then obtain
\be
Q^2={k_1\over c_0^2Z_0}\sp L^2={k_2\over k_1}{Z_0c_0^2\over W_0}
\label{78o}\ee
and
\be
k_1k_3=\left(z+{a_0\e\over 2}\right)^2k_2{Z_0V_0\over W_0}
\label{79o}\ee

\subsubsection{Option 1}

With option 1 as in  (\ref{14o})
we obtain an extra relation
\be
k_3={Z_0V_0\over 4W_0}(2z+a_0\e)(a_0\e+4-2a_0\g)
\label{80o}\ee
We must now solve the two remaining equations (\ref{79o}) and (\ref{80o}).
We can derive from the two
\be
(a_0\e+4-2a_0\g)k_1=(2z+a_0\e)k_2
\label{81o}\ee
This equation has two solutions.

\begin{enumerate}

\item\be
a_0={(\g-\d)z-(\g+\d)\over 1-\d^2}
\label{85o}\ee

and the leftover condition for $z$
becomes
\be
\text{W}_0= -\frac{\text{V}_0 \text{Z}_0 \left(2  \delta ^2(z^2- z+ \theta) +(1-\delta ^2) \theta ^2\right) \left(2\delta ^2(2-2 z+ z \theta) +(1-\delta ^2) \theta ^2\right)}{2 (-1+z)^2 \delta ^2 \left(8 \delta ^2+2 z \delta ^2+2 z^2 \delta ^2-6 \delta ^2 \theta -2 z \delta ^2 \theta +\theta ^2+\delta ^2 \theta ^2\right)}
\label{86o}\ee
where we used
\be
	\gamma =\frac{\delta ^2(1+z)+\theta (1-\delta ^2) }{(-1+z) \delta }\,.
\ee
The other parameters of the solution read
\be
	\begin{split}
	&a_0=\frac{\theta}{\delta}\,,\quad L^2= \frac{8 \delta ^2+2 z \delta ^2+2 z^2 \delta ^2-6 \delta ^2 \theta -2 z \delta ^2 \theta +\theta ^2+\delta ^2 \theta ^2}{2V_0 \delta ^2}\\
	&c_0=\frac{2 \delta ^2(z- z^2- \theta)+(\delta ^2-1) \theta ^2}{2 (-1+z) \delta ^2}\,, \quad Q^2= \frac{-2 \delta ^2 (-1+z)^2}{Z_0(2 \delta ^2(z- z^2- \theta)+(\delta ^2-1) \theta ^2)}
	\end{split}
\ee
This is the solution described in section \ref{section:massiverunning}.

\item The other solution is
\be
\theta=z+2
\label{82o}\ee
and from (\ref{80o}) we obtain a quadratic equation for $z$
\be
W_0=-\frac{V_0 Z_0(2 \gamma +z \gamma -2 \delta +z \delta )^2}{2 \left(4+4 z+z^2-4 z \delta ^2+z^2 \delta ^2\right)}
\label{83o}\ee
and the rest of the solution reads
\be
\begin{split}
&a_0=\frac{\theta}\d\,,\quad L^2= \frac{4+4 z+z^2-4 z \delta ^2+z^2 \delta ^2}{2V_0 \delta ^2} \\
&c_0= -\frac{2 \gamma +z \gamma -2 \delta +z \delta }{2 \delta }\,,\quad Q^2= \frac{2 (4+4 z (1-\delta^2 )+z^2 \left(1+\delta ^2\right))}{Z_0 (2 \gamma +z \gamma -2 \delta +z \delta )^2}
\end{split}
\ee
Note that requiring $r$ to be spacelike ($L^2>0$) and $Q^2>0$ implies $W_0<0$, which is ill-defined, as it is ultimately related to the square of the charge of the original complex scalar field.

\end{enumerate}

\subsubsection{Option 2}

There is no consistent solution to the equations of motion: a solution to Maxwell and Einstein's equations does not solve the scalar equation.

\subsection{$k_3=0$}

From the equations (\ref{4o}) we have
\be
c^2Q^2Z_0=k_1\sp L^2Q^2W_0=k_2
\label{89o}\ee
\be
 c_0=-z-{a_0\e\over 2}\sp \t=a_0(\g-\e)
\label{90o}\ee
as well as
\be
k_3=
a_0^2+2(z^2+z+4)-2\t(z+3)+\t^2=0
\label{91o}\ee

\subsubsection{Option 1}

 With option 1  we obtain also
\be
L^2=c_0(c_0+a_0\g+z-2){Z_0\over W_0}
\label{92o}\ee
We end up with
\be
Q^2={k_1\over c_0^2Z_0}\sp L^2={k_2\over k_1}{Z_0c_0^2\over W_0}\sp  c_0=-z-{a_0\e\over 2}\sp \t=a_0(\g-\e)
\label{93o}\ee
and
\be
\begin{split}
&\left(2+z +(\epsilon-\gamma)  a_0\right) \left(-2 \gamma +\epsilon +z \epsilon +a_0((\gamma-\epsilon)^2-1)\right)=0\,,\\
&4+z+z^2-(3+z) \theta +\frac{\theta ^2}{2}+\frac{a_0^2}{2}=0\,.
\end{split}
\label{94o}\ee
Like in the previous case the first equation has two solutions for $z$.

\begin{enumerate}

\item The first solution is
\be
a_0=\frac{2 \gamma -\epsilon -z \epsilon }{(-1+\gamma -\epsilon ) (1+\gamma -\epsilon )}\,,\quad \theta=\frac{(2 \gamma -\epsilon -z \epsilon)(\g-\d) }{(-1+\gamma -\epsilon ) (1+\gamma -\epsilon )}\
\label{96o}\ee
while the second equality in \eqref{94o}  becomes a quadratic equation in $z$.
The problem with this solution is that imposing consistency constraints $L^2>0$, $Q^2>0$ and $a_0^2>0$ is inconsistent with $(\theta-2)(\theta-2z)>0$ (well-defined UV).

\item The other solution is
\be
\theta=2+z\,,\quad a_0=\frac{2+z}{\g-\e}
\label{95o}\ee
and the first equation in (\ref{94o}) becomes the quadratic equation for $z$
\be
2(\gamma -\epsilon )^2-2 z (-1+\gamma -\epsilon ) (1+\gamma -\epsilon )+z^2\left(1+\gamma ^2-2 \gamma  \epsilon +\epsilon ^2\right)=0
\label{97o}\ee
that has real roots if $(\e-\g)^2\geq 3$.
It necessarily has either $L^2<0$ or $W_0<0$, so we discard it.

\end{enumerate}

\subsubsection{Option 2}

In that case, one can find a leading order solution:
\be
	\begin{split}
	&L^2=-\frac{Z_0 (-2-z+\theta )^2 (-1-z+\theta )}{W_0 (-2+\theta )}\,,\quad a_0=\sqrt{-8-2 z-2 z^2+6 \theta +2 z \theta -\theta ^2}\,,\\
	&\e^2=\frac{4 (2-\theta )}{8+2 z+2 z^2-6 \theta -2 z \theta +\theta ^2},\quad \g^2=\frac{-(-4+\theta )^2}{8+2 z+2 z^2-6 \theta -2 z \theta +\theta ^2},\\
	&Q=\sqrt{\frac{2(2-\theta)}{Z_0(-2-z+\theta)}}\,, \quad c_0=-2-z+\theta\,,
	\end{split}
\ee
but the subleading series is logarithmic.

\subsection{$k_1=0$}

From (\ref{8o}) we have
\be
a^2+2z(z-1)-\t(\t-2)=0
\label{100o}\ee
and since $k_2,k_3\not=0$

\be
L^2V_0=k_3\sp L^2Q^2W_0=k_2
\label{101o}\ee
\be
 c_0=-z-{a_0\e\over 2}\sp \t=a_0\d
\label{102o}\ee
so that
\be
Q^2={k_2\over k_3}{V_0\over W_0}\sp L^2={k_3\over V_0}
\label{103o}\ee
There remains to impose the gauge field equation.

\subsubsection{Option 1}

There is no consistent solution to the equations of motion: a solution to Maxwell and Einstein's equations does not solve the scalar equation.

\subsubsection{Option 2}

There is only one consistent leading order solution
\be
	\begin{split}
	& L^2=\frac{(-2+\theta ) (-2-z+\theta )}{V_0}\,, a_0=\pm\sqrt{2 z-2 z^2-2 \theta +\theta ^2}\,,\\
	&\e=\mp\frac{2 z}{\sqrt{2 z-2 z^2-2 \theta +\theta ^2}},\quad \d=\pm\frac{\theta }{\sqrt{2 z-2 z^2-2 \theta +\theta ^2}},\\
	&Q=\sqrt{\frac{2V_0(-1+z)}{W_0(2-\theta)}}\,,\quad c_0=0\,.
	\end{split}
\ee
However, the next order in the power series is inconsistent: the solution to Maxwell and Einstein's equations is not a solution to the scalar equation.

\subsection{$k_2=0$}

\subsubsection{Option 1}

There is no consistent solution to the equations of motion: a solution to Maxwell and Einstein's equations does not solve the scalar equation.

\subsubsection{Option 2}

There is a single leading solution, which as expected is a correction to the massless case \eqref{MasslessSolEMD}:
\be
	\begin{split}
	& L^2=\frac{(-2-z+\theta ) (-1-z+\theta )}{V_0}\,,\quad Q=\sqrt{\frac{2(-1+z)}{Z_0 (2+z-\theta ) }}\\
	&\gamma = \mp\frac{-4+\theta }{\sqrt{-(-2+2 z-\theta ) (-2+\theta )}},\quad\delta = \pm\frac{\theta }{\sqrt{(-2+\theta ) (2-2 z+\theta )}}\\
	&a_0 = \pm\sqrt{(-2+\theta ) (2-2 z+\theta )},\quad c_0=-2-z+\theta\,.
	\end{split}
\ee
The series expansion has $\a=4 (-\gamma +\delta +\epsilon )/(\gamma +\delta )$. All amplitudes are proportional to the ratio $W_0/(V_0Z_0)$.%:

\subsection{$k_1=k_2=0$\label{appC.5}}

\subsubsection{Option 1}

There are two leading order solutions, where in both cases the metric and the scalar are the same as in the neutral domain-wall \eqref{neutralDW2} and one has to set $\epsilon=\gamma-\delta$:
\be
	z=1\,,\quad L^2=\frac{(-3+\theta ) (-2+\theta )}{V_0}\,,\quad a_0= \sqrt{\theta (\theta -2)},\quad \theta = \frac{2 \delta ^2}{(-1+\delta^2 )}\ee
They differ by the gauge field:
\begin{enumerate}
	\item  $Q=0$. One has zero leading order gauge field and so solves exactly the equations of motion, without a power series.
	\item The other verifies
	\be
		W_0=-\frac{V_0Z_0c_0^2}{(-3+\theta ) (-2+\theta )},\quad \gamma = -\frac{-1+2 c_0}{\sqrt{(-2+\theta ) \theta }}\,,
	\ee
	from which one determines the scaling of the gauge field, while the charge $Q$ is independent from the other variables. However, the power series scales like $r^{3-\theta}$, and this leads to a contradiction: this is the same scaling as the temperature fluctuation, which should be relevant in the IR and not irrelevant.
\end{enumerate}

\subsubsection{Option 2}

There are three leading order solutions, where in all cases the metric and the scalar are the same as in the neutral domain-wall \eqref{neutralDW2}
\be
	z=1\,,\quad L^2=\frac{(-3+\theta ) (-2+\theta )}{V_0}\,,\quad a_0= \sqrt{\theta (\theta -2)},\quad \theta = \frac{2 \delta ^2}{(-1+\delta^2 )}\ee
They differ by the gauge field:
\begin{enumerate}
	\item $Q=0$. One has zero leading order gauge field and so solves exactly the equations of motion, without a power series. Linear perturbations around this solutions are all relevant, so we discard it.
	\item The other has constant gauge field $c_0=0$. It has a consistent power series with a different power for the gauge field and for the other fields
	\be
		\a_1=(\epsilon -\gamma)  \sqrt{(\theta-2)\theta }+\theta\,\quad	\a_2=2+\epsilon   \sqrt{(\theta-2)\theta }
	\ee
and
\be
	\begin{split}
	&\mathfrak a_1=\frac{Q L^2W_0}{ Z_0 c_1 \left(c_1-1+\gamma  \sqrt{\theta(\theta-2)}\right)}, \quad\mathfrak b_1= \frac{Q^2L^2W_0 (\theta-2)}{ \beta  (-3+\beta +\theta )},\quad \mathfrak c_1=0\,,\\
	& \mathfrak d_1= \frac{Q^2L^2 W_0}{\beta  (-3+\beta +\theta )},\quad \varphi _1= -\frac{Q^2L^2W_0 (-2+\beta )}{2  \beta  \sqrt{-2+\theta } \sqrt{\theta } (-3+\beta +\theta )}\,.
	\end{split}
\ee
$Q$ is an integration constant left undetermined by the field equations.
	\item The third has a non-constant gauge field at leading order but yields a logarithmic series.
\end{enumerate}

\subsection{$k_1=k_3=0$ }

\subsubsection{Option 1}

There is no consistent solution to the equations of motion: a solution to Maxwell and Einstein's equations does not solve the scalar equation.

\subsubsection{Option 2}

In that case there is no charged solution with a running scalar.

\subsection{$k_2=k_3=0$}

\subsubsection{Option 1}
There is no consistent solution to the equations of motion: a solution to Maxwell and Einstein's equations does not solve the scalar equation.

\subsubsection{Option 2}

The solution is
\be
	a_0= \sqrt{(3-z)(z-1)},\quad \theta = 1+z\,,\quad Q=\sqrt{\frac{2(z-1)}{Z_0}}, \quad c_0=-1,\quad \gamma = \sqrt{\frac{3-z}{z-1}}
\ee
Note that $L^2$ is left undetermined at leading order, which is pathological.

\subsection{$k_1=k_2=k_3=0$}

\subsubsection{Option 1}

The solution is
\be
	\begin{split}
	&a_0= \sqrt{4 z-z^2},\quad \theta = 2+z\,,\quad \epsilon = \frac{-2-z+\sqrt{(4-z) z} \gamma }{\sqrt{(4-z) z}},\\
	&W_0= \frac{Z_0 c_0 \left(z-2+\sqrt{(4-z) z} \gamma +c_0\right)}{L^2}
	\end{split}
\ee
but the subleading series is logarithmic.

\subsubsection{Option 2}

In this case, there are three solutions, with identical metric and scalar at leading order:
\be
	a_0= \sqrt{4 z-z^2},\quad \theta = 2+z
\ee
and differing only by the value taken by the gauge field. However, the equations of motion leave $L^2$ free in all cases, so we discard them all.
\begin{enumerate}
\item The first has zero gauge field at leading order ($Q=0$).
\item The second has constant gauge field at leading order ($c_0=0$) while $Q$ is a free parameter.
	\item The third solution has a non-constant gauge field at leading order but only allows for a logarithmic series.
\end{enumerate}

\section{Linear perturbations\label{app:LinPert}}

In this appendix, we give details about the linear perturbations around the various (leading order) solutions we have analyzed. They are of two types: either they are exact solutions to the field equations, or these leading order solutions have to be supplemented by a power series. There is a slight subtlety with the latter: they are an expansion in the radial coordinate (powers become subleading in the IR), while the linear perturbations are an expansion with a small amplitude. So, to leading order both in $r$ and in the amplitudes of perturbations, it is enough to work out the linear perturbation around the leading order solution, discarding the power series. In this fashion, linear perturbations around the solution in section \ref{section:masslessrunningbr} are identical to those around the solution in section \ref{section:masslessrunning}. The two solutions only differ by the power series in section \ref{section:masslessrunningbr}, which reflects the fact we have allowed $W(\phi)\neq0$ there.

We perturb around the solution \eqref{2o} by setting
\be \label{perturbationsconstant}
\begin{split}
	 B=&L^2r^{\theta-2}\left(1+B_1r^{b_1}\right),\quad D=r^{\theta-2z}\left(1+D_1r^{d_1}\right),\quad C=r^{\theta-2}\left(1+C_1r^{c_2}\right),\\
	\phi=&\phi_0+ \phi_1 r^{a_1},\quad A=r^{c_0}\left(Q+A_1r^{c_1}\right).
\end{split}
\ee
in the case where the scalar is a constant at leading order (without any warping factor in $C(r)$ in the AdS$_2\times R^2$ case), or
\be \label{perturbations}
\begin{split}
	B=&L^2r^{\theta-2}\left(1+B_1r^{b_1}\right),\quad D=r^{\theta-2z}\left(1+D_1r^{d_1}\right),\quad C=r^{\theta-2}\left(1+C_1r^{c_2}\right),\\
	\phi=&a_0\log r+ \phi_1 r^{a_1},\quad A=r^{c_0}\left(Q+A_1r^{c_1}\right).
\end{split}
\ee
if it is running.
Assuming the IR to be at $r\to+\infty$ ($r\to0$), a typical mode $\beta$ will be \emph{irrelevant in the IR} if it is negative (positive). It is also fruitful to distinguish between ``universal'' modes, which depend only on the exponents $z$ and $\theta$ in the metric, and ``non-universal'' modes, which depends on the details of the theory, \cite{GubserNellore}, as well as between those which drive a flow to finite temperature and those that allow to interpolate between different ground states. For this, we evaluate the extremality charge $\mathcal{Q}$, \eqref{extrCharge}.

We are mostly interested in zero-temperature RG flows, where a scale invariant fixed point mediates a fractionalization transition between two hyperscaling violating quantum critical lines. For the former geometries, we will study whether one of the zero-temperature deformations can be relevant, so that the fixed point is unstable. For the latter, we will require that only irrelevant deformations exist so that the RG flow is well-defined (most of the time it needs to be constructed numerically). Indeed, the perturbations will typically generate a two-parameter family of solutions. One parameter can be used to satisfy the UV boundary condition on the condensing scalar: the source to its dual operator should be set to zero on the boundary, so that the U(1) symmetry is spontaneously broken and the boundary theory is in a superfluid state. This leaves generically a one-parameter family of superfluid phases.

\subsection{AdS$_4$ geometry \label{app:AdS4}}

 We perturb the solution \eqref{AdSDW} using the Ansatz \eqref{perturbations}. There is a universal mode $b_1=d_1=\b_u=3$ which is simply the temperature perturbation with $D_1=-B_1$ and the rest set to zero (and indeed $\mathcal Q\neq0$). We know that this matches the analytical finite temperature completion, so we do not have to worry that this perturbation could source the scalar or the gauge field at higher orders. All the other perturbations satisfy $\mathcal Q=0$ and so maintain zero temperature. The conjugate mode with $d_1=0$ and $D_1\neq0$ is a time rescaling.

There is a scalar perturbation $a_1=\b^\phi$ with $\phi_1\neq0$, which is a solution of
\be
	0=\left(\b^\phi\right)^2 -3\b^\phi +L^2 V''_\star\,,\qquad 9-4L^2 V''_\star>0\,.
\ee
 Note that it displays the usual instability when the scalar mass goes below the Breitenlohner-Friedman bound, $V''_{BF}=9/4L^2<0$. Below that bound, the dimension of the operator dual to that mode becomes complex, and one expects another IR geometry to develop (such as Lifshitz, \cite{GubserNellore, Horowitz:2009ij}). The two modes read
\be
	\b^\phi_\pm=\frac12\left[3\pm\sqrt{9-4L^2V''_\star}\right]
\ee
with $\b_-^\phi<0$ providing the irrelevant perturbation we need.

The second irrelevant mode is provided by the gauge field perturbations with $c_1=\b^q$ and $A_1\neq0$. They decouple from the other perturbations, as the charge dynamics do not impact the metric field equations once the scalar is frozen to a non-zero value. One finds:
\be
0=-6 W_\star-\b^q V_\star Z_\star+\left(\b^q\right)^2 V_\star Z_\star\,,\qquad 24 W_\star+V_\star Z_\star>0\,.
\ee
As we are interested in approaching AdS$_4$ in the IR, we should select the irrelevant mode:
\be
	\b^q_\pm=\frac12\left[1\pm\sqrt{1+\frac{4L^2 W_\star}{ Z_\star}}\right].
\ee
$\b^q_+>0$ while $\b^q_-<0$, so the latter can be used to drive the flow to the IR AdS$_4$. In the case where the U(1) is unbroken $W_\star=0$, then the modes reduce to $\b^q_\mp=0,1$ as usual for a UV AdS$_4$ (chemical potential and charge density).

\subsection{AdS$_2\times R^2$ geometry\label{app:AdS2R2}}

Perturbing around the solution \eqref{AdS2xR2br} using the Ansatz \eqref{perturbationsconstant}, one finds a universal, relevant mode with power $\beta=1$ and amplitudes:
\be
\begin{split}
&C_1= -\frac{\phi _1 W'_\star}{2 V_\star Z_\star},\qquad D_1= -B_1-\frac{\phi _1 \left(2 Z_\star V'_\star+W'_\star\right)}{V_\star Z_\star}\,,\qquad A_1=0\,.
\end{split}
\ee
 When $W(\phi)=0$, this is just the finite temperature completion of AdS$_2\times R^2$ (only $D_1$ is non-zero). In the massive case, we see that the perturbation now sources both the scalar ($\phi_1$) and the metric components of the $R^2$ factor ($C_1$). This perturbation generically has $\mathcal Q\neq0$, but one can choose a linear combination of $\phi_1$ and $B_1$ which maintains zero temperature.
The two conjugate modes $\b=0$ with $D_1= 2 A_1$ and $C_1$ free are respectively a rescaling of time and of the volume of the spatial directions.

Then, we also find four other modes, which drive zero temperature flows ($\mathcal Q=0$), and might be real or complex depending on the value taken by the coupling functions and their first/second derivatives at $\phi=\phi_\star$. By tuning these, one may find both relevant and irrelevant modes. They satisfy a fourth-order polynomial
\be
	(\b-\b_+^1)(\b-\b_-^1)(\b-\b_+^2)(\b-\b_-^2)=0\quad \Rightarrow \quad\b_\pm^i=\frac12\left(1\pm\sqrt{X\pm\sqrt{Y}}\right),
\ee
with
\be
	\begin{split}
	&\sum\beta _\pm^i=2\,,\quad \sum_{i<j}\b _\pm^i\b _\pm^j=-1-\frac{2 Z'_\star{}^2}{Z_\star{}^2}+\frac{V''_\star}{V_\star}+\frac{W''_\star}{V_\star Z_\star}+\frac{Z''_\star}{Z_\star}=\frac{3-X}2\,,\\
	&\sum_{i<j<k}\b _\pm^i\b _\pm^j\b _\pm^k=\sum_{i<j}\b _\pm^i\b _\pm^j-1=\frac{1-X}2\,,\\
	&\prod\b _\pm^i=-\frac{V'_\star{}^2}{V_\star{}^2}+\frac{4 V'_\star Z'_\star}{V_\star Z_\star}+\frac{Z'_\star{}^2}{Z_\star{}^2}+\frac{2 V''_\star}{V_\star}+\frac{2 W''_\star}{V_\star Z_\star}+\frac{2 Z''_\star}{Z_\star}=\frac{(1-X)^2-Y}{16}\,.
	\end{split}
\ee
The modes go by pairs, summing to $1$, so that if they are complex, their reals parts are all positive, equal to $1/2$ and the fixed point is dynamically unstable.
Irrespective of the precise value taken by $\b$, the amplitudes read (after gauging away $A_1$):
\be
\begin{split}
&B_1= -\frac{\phi _1 \left(2 \left(-1+\beta ^2\right) Z_\star V'_\star+\left(1-2 \beta +2 \beta ^2\right) W'_\star\right)}{(-2+\beta ) (1+\beta ) V_\star Z_\star}\,,\\
	& D_1= \frac{\phi _1 \left(2 (1+\beta ) Z_\star V'_\star+(1+2 \beta ) W'_\star\right)}{(-2+\beta ) (1+\beta ) V_\star Z_\star}\,,\qquad C_1= -\frac{\phi _1 W'_\star}{\beta  (1+\beta ) V_\star Z_\star}\,.
\end{split}
\ee

\subsection{Lifshitz geometry\label{app:Lifshitz}}

Perturbing around the solution \eqref{MassiveSolConstantPhi} using the Ansatz \eqref{perturbationsconstant}, there exist two universal modes: one is irrelevant $\b=0$ and corresponds to a rescaling of time; one is relevant, $\beta_u=2+z$, and puts the solution at finite temperature without turning on the scalar; as well as two pairs of conjugate modes. Each pair sums to $2+z$: two modes will have a fixed sign and be relevant, while typically the other two can be (ir)relevant depending on parameters. Their value is determined from the quartic polynomial:
\be
	(\b-\b_+^1)(\b-\b_-^1)(\b-\b_+^2)(\b-\b_-^2)=0\quad \Rightarrow \quad\b_\pm^i=\frac12\left(2+z\pm\sqrt{X\pm\sqrt{Y}}\right),
\ee
with
\be
	\begin{split}
	&\sum\beta _\pm^i=2(2+z)\,,\\
	& \sum_{i<j}\b_\pm^i\b_\pm^j=(10-z) z+\frac{\left(4+z+z^2\right) }{V_\star}\left(V''_\star+\frac{(z-1) W''_\star}{zZ_\star}\right)+(z-1) z\left(\frac{Z''_\star}{Z_\star}-2\frac{Z'_\star{}^2}{Z_\star^2}\right)\\
	&\qquad\qquad=\frac{1}{2}\left(12+12 z+3 z^2-X \right)\,,\\
	&\sum_{i<j<k}\b_\pm^i\b_\pm^j\b_\pm^k= \left(z+2\right)^3-(z+2) \sum_{i<j}\b_i\b_j=\frac{1}{2} (2+z) \left(4+4 z+z^2-X \right)\,,\\
	&\prod\b_\pm^i=2(z-1)(z-2) \sum_{i<j}\b_\pm^i\b_\pm^j-2 (-10+z) (-2+z) (-1+z) z\\
	&+\frac{(1+z) \left(4+z+z^2\right)^2 V'_\star{}^2}{(-1+z) V_\star{}^2}-\frac{2 z (-5+2 z) \left(4+z+z^2\right) V'_\star Z'_\star}{V_\star Z_\star}-\frac{5 (-1+z)^2 z^2 Z'_\star{}^2}{Z_\star{}^2}\\
	&\qquad=\frac{1}{16} \left(4+4 z+z^2-X \right)^2-\frac{Y }{16}\,.
	\end{split}
\ee
In all cases, one can express the amplitudes of the perturbation in terms of $\phi_1$. The expressions are not particularly enlightening and very lengthy, so we do not give them.

\subsection{Neutral geometry with hyperscaling violation\label{app:NeutralHypVio}}

Perturbing around the solution \eqref{neutralDW2} with the Ansatz \eqref{perturbations}, with a relevant, universal mode $d_1=b_1=a_1=\beta_u=3-\theta$:
\be
	\begin{split}
	&\phi_1=\sqrt{\theta (\theta -2)}\tilde \phi _1\,,\quad C_1 =0\,,\quad D_1=-B_1+2\theta \tilde\phi _1\,,
	\end{split}
\ee
while $A_1$ decouples from the others.
There are two free amplitudes, $B_1$ and $\phi_1$, while the extremality charge $\mathcal Q$ is proportional to $D_1$. So any flow with $D_1\neq0$ will go to finite temperature (including the case $\phi_1=0$, for which there is an exact analytical finite temperature completion, \cite{cgkkm,gk}). On the other hand, imposing $D_1$ picks up  a zero temperature flow.
One also finds two modes conjugate to the previous two, with $d_1=b_1=a_1=0$, $B_1=\d\phi_1$ and $D_1$ free. The former is  an infinitesimal, constant shift of $\phi$, the latter a time reparametrization.

Turning to the gauge field perturbations (and using \eqref{EMDmassiveCouplings}), $c_1=\b^q$, there exists a pair of conjugate modes:
\be
\begin{split}
	&B_1=C_1=D_1=\phi_1=0\,,\quad A_1\neq0\,,\quad \e=\g-\d\,,\\
	 &\beta^q_\pm=\frac12\left(1-\sqrt{\theta(\theta-2)}\g\pm\sqrt{\left(1-\sqrt{\theta(\theta-2)}\g\right)^2+4L^2\frac{W_0}{Z_0}}\right),\end{split}
\ee
both sourced by $A_1$.
$\beta^q_-<0$, so it is always an irrelevant perturbation and it can drive a $T=0$ flow from the UV to the neutral dilatonic fixed point.

For the power series solution, \eqref{neutralDWPowerSeries}, the gauge field perturbations are slightly modified to
\be
\beta^q_+=1-\sqrt{\theta(\theta-2)}\g\,,\qquad \b^q_-=0\,,
\ee
since terms in $W_0$ now become subleading. Moreover, the constraint $\e=\g-\d$ is evaded.

\subsection{Charged, fractionalized geometry with hyperscaling violation\label{app:ChargedFracHypVio}}

Perturbing the solution \eqref{MasslessSolEMDPower} linearly  with the Ansatz \eqref{perturbations}, there are two, doubly-degenerate universal modes  (independent on the details of the scalar couplings): one is irrelevant $\b=0$
\be
	\phi_1=\sqrt{(\theta-2)(2-2z+\theta)}\,\tilde\phi_1\,,\quad B_1=\theta\tilde\phi_1\,,\quad D_1=(4-\theta)\tilde\phi_1\,,
\ee
and is just a constant, infinitesimal shift of the scalar, while the other has $A_1=2D_1$ and is just a time rescaling. The conjugate universal mode is relevant, $\beta=\beta_u=2+z-\theta$:
\be
	\phi_1 =0\,,\qquad D_1=-B_1\,,\qquad A_1\neq0
\ee
with $B_1$ controlling the strength of the perturbation; $A_1$ decouples and should simply be identified with the chemical potential. Then, $B_1$ generates temperature ($\mathcal Q\neq0$), as can be verified by putting the solution at finite temperature exactly, \cite{cgkkm}. Choosing a different gauge ($C_1\neq0$), one may pick a linear combination of the amplitudes maintaining zero temperature.

The two remaining modes $\beta_\pm$  are non-universal, and always maintain zero temperature ($\mathcal Q=0$). Their sum is also $2+z-\theta$, with typically one of them irrelevant and the other relevant:
\be
	\begin{split}
	&\phi_1=\sqrt{(\theta-2)(2-2z+\theta)}\,\tilde\phi_1\,,\quad B_1= (2-2z+\theta)\tilde\phi_1\,,\\
	& D_1=(2-2z+\theta)\tilde\phi_1,\quad A_1= \frac{2-2z+\theta}{2(1-z)\beta_\pm}\tilde\phi_1\,,\\
	&\beta_\pm=\frac12\left(2+z-\theta\pm\sqrt{\frac{2+z-\theta}{2z-2-\theta}\left(-20+2 z+18 z^2+16 \theta -19 z \theta +\theta ^2\right)}\right).
	\end{split}
\ee

\subsection{Charged, cohesive geometry with hyperscaling violation\label{app:ChargedCohHypVio}}

 We linearly perturb the solution \eqref{MassiveSolRunningScalar}, using \eqref{perturbations}. One first finds the universal, relevant pertubation $\beta=\beta_u=2+z-\theta$:
\be
\begin{split}
A_1=& \frac{\left(-\theta ^2+\delta ^2 \left(-4-2 z^2+2 \theta +z \theta \right)\right) B_1}{2 (-1+z) \delta ^2 (2+z-\theta )}\\
&+\frac{\left((2+z) \theta ^3+\delta ^4 (\theta-2 ) \left(2 z^3-24-10 z^2+20 \theta -6 \theta ^2+\theta ^3-2 z \left(2-5 \theta +\theta ^2\right)\right)\right)\phi_1}{2 (-1+z) \delta  (2+z-\theta ) \left(\delta ^2 (-2+2 z-\theta ) (-2+\theta )+\theta ^2\right)}\\
&+\frac{\left(\delta ^2 \theta  \left(16+2 z^3+2 z^2 (3-\theta )-8 \theta +2 \theta ^2-\theta ^3+z \left(12-14 \theta +3 \theta ^2\right)\right)\right) \phi _1}{2 (-1+z) \delta  (2+z-\theta ) \left(\delta ^2 (-2+2 z-\theta ) (-2+\theta )+\theta ^2\right)}\,,\\
D_1=& -\frac{\left(2 \delta ^2-3 z \delta ^2+z^2 \delta ^2+\delta ^2 \theta +z \delta ^2 \theta +\theta ^2-\delta ^2 \theta ^2\right) B_1}{(-1+z) \delta ^2 (2+z-\theta )}\\
&+\frac{\left(8 \delta ^2+2 z \delta ^2+2 z^2 \delta ^2+4 \theta +2 z \theta -6 \delta ^2 \theta -2 z \delta ^2 \theta -\theta ^2+\delta ^2 \theta ^2\right) \phi _1}{(-1+z) \delta  (2+z-\theta )}\,.
\end{split}
\ee
The mode is doubly-degenerate: two amplitudes (for instance $B_1$, $\phi_1$) remain independent and can drive separate relevant flows, typically to finite temperature ($\mathcal Q\neq0$, but one can pick a gauge maintaining zero temperature). Comparing with the Lifshitz case (section \ref{section:Lifshitz}) where we had four conjugate non-universal modes, two have collapsed to universal values upon choosing couplings such as \eqref{EMDmassiveCouplings}. Consequently, there is also a doubly-degenerate, irrelevant $\b=0$ mode (always with $\mathcal Q=0$): the two independent amplitudes can be chosen to be $A_1$ and $\phi_1$. The $A_1$ perturbation is just a rescaling of the time coordinate, while the other is a constant shift of the scalar.

There is also a pair of conjugate modes $\b_++\b_-=2+z-\theta$, one relevant and one irrelevant in the IR,  solutions to
\be
\begin{split}
&\left(-2 \delta ^2+3 z \delta ^2-z^2 \delta ^2-\delta ^2 \theta -z \delta ^2 \theta -\theta ^2+\delta ^2 \theta ^2\right) \times\\
&\left(2 \delta ^2-4 z \delta ^2+2 z^2 \delta ^2+2 \delta ^4+2 z \delta ^4+3 \delta ^2 \theta +z \delta ^2 \theta -3 \delta ^4 \theta -z \delta ^4 \theta +\theta ^2-2 \delta ^2 \theta ^2+\delta ^4 \theta ^2\right)\\
&-(-1+z)^2 \delta ^4 (2+z-\theta ) \b+(-1+z)^2 \delta ^4 \b^2=0\,.
\end{split}
\ee
One can choose to express $A_1$, $B_1$ and $D_1$ in terms of $\phi_1$, while setting $C_1=0$:
\be\begin{split}
A_1=&-\frac{\delta  \left(-2 z \delta ^2+2 z^2 \delta ^2+\theta  \left(-\delta ^2 (-2+\theta )+\theta \right)\right)\beta_\pm \phi _1}{2 \left(\delta ^2 (-2+\theta )-\theta \right) \left(\theta ^2+\delta ^2 \left(2+z^2+z (-3+\theta )+\theta -\theta ^2\right)\right)}\\
B_1=&\frac{\left(\theta ^2+\delta ^2 \left(2-4 z+2 z^2+2 \theta -\theta ^2\right)\right) \phi _1}{\delta  (1+z-\theta ) \left(\delta ^2 (-2+\theta )-\theta \right)},\\
&+\frac{(1-z) \delta  \left(\delta ^2 (-2+2 z-\theta ) (-2+\theta )+\theta ^2\right)\beta_\pm \phi _1}{(1+z-\theta ) \left(\delta ^2 (-2+\theta )-\theta \right) \left(\theta ^2+\delta ^2 \left(2+z^2+z (-3+\theta )+\theta -\theta ^2\right)\right)}\\
D_1=& \frac{2 (-1+z) \delta  \phi _1}{\delta ^2 (-2+\theta )-\theta }\,.
\end{split}
\ee
For these modes, $\mathcal Q=0$, so the flows are at $T=0$.

\section{Kaluza-Klein compactifications\label{app:DimRed}}

There are two straightforward ways of getting the action \eqref{1} with scalar potential and gauge couplings given by \eqref{EMDmassiveCouplings}: either by starting from an action containing a massive vector field or a massive higher-rank form (reduced to a lower-dimensional vector field). We will deal with these two possibilities in turn, and also eventually add a higher-dimensional scalar. We will focus on toroidal reductions, but it is straightforward to extend the calculations below to a reduction along a sphere. Throughout this section, higher-dimensional quantities will be denoted by a tilde and have latin capital indices, as in $\tilde g_{AB}$ for the $(p+n+1)$-dimensional metric. On the other hand, untilded quantities will be lower dimensional and have small greek indices, such as $g_{\mu\nu}$ for the $(p+1)$-dimensional metric. We give results for arbitrary dimension $p$.

\subsection{From a higher-dimensional massive vector field\label{app:C.1}}

We start from the higher-dimensional action
\be
\tilde  S=\tilde M^{2}\int d^{p+n+1}x\sqrt{-\tilde g}\left[\tilde R+2\Lambda-\frac14\left(\tilde F_{[2]}\right)^2-\frac12\left(\tilde A_{[1]}\right)^2\right]
\label{4.1.1}
\ee
where $F_{[2]}=\ud A_{[1]}$ is a two-form field strength for a Maxwell potential, $\tilde g$ the $(p+n+1)$-dimensional metric and $\left(\tilde F_{[2]}\right)^2=\tilde g^{AC}\tilde g^{BD}F_{[2]AB}F_{[2]BD}$ and $\left(\tilde A_{[1]}\right)^2=\tilde g^{BC}A_{[1]B}A_{[1]C}$.

Then, using the reduction Ansatz
\be
	\ud \tilde s^2 = e^{-\da\phi}\ud s^2+ e^{\frac{\phi}{\da}\left(\frac2{p-1}-\da^2\right)}\ud R^2_{(n)} \,,\label{4.1.2}
\ee
for the metric with
\be
	\da^2={2\over p-1}~\frac{n}{(p+n-1)}\leq \da_c^2={2\over p-1}\,,
\label{4.1.3}\ee
and assuming the gauge potential to depend only on the lower-dimensional coordinates $x^\mu$ and without any legs along the reduced dimensions, one obtains for the gauge field quadratic invariants
\be
	\left(\tilde F_{[2]}\right)^2 = e^{2\da\phi}\left(F_{[2]}\right)^2,\qquad \left(\tilde A_{[1]}\right)^2=e^{\da\phi}\left(A_{[1]}\right)^2. \label{4.1.4}
\ee
Combining with
\be
	\sqrt{-\tilde g}=\sqrt{-g}e^{-\da\phi}, \label{4.1.5}
\ee
the reduced action is
\be
 S=M^{2}\int d^{p+1}x\sqrt{- g}\left[ R-\frac12\partial\phi^2+2\Lambda e^{-\da\phi}-\frac14e^{\da\phi}\left( F_{[2]}\right)^2-\frac12\left( A_{[1]}\right)^2\right]
\label{4.1.6}
\ee
which has
\be
	\ga=\da\,,\qquad \epsilon =0\,,\qquad \ga-\da=\epsilon \label{4.1.7}
\ee
and so verifies the relations $\ga-\da=\epsilon$ for any dimension $p$ and in particular for $p=3$.

Setting $\ga=\da$, $\epsilon=0$ in \eqref{MassiveSolRunningScalar}, and using \eqref{4.1.3}, the higher-dimensional solution reads, after rescaling the coordinates,
\bea
	\ud \tilde s^2&=&-{\ud t^2\over u^{2z}}+L^2{\ud u^2\over u^2}+{\ud x^2+\ud y^2+\ud R^2_{(d-3)}\over u^2},\label{4.1.10}\\
	A_{[1]}&=&\sqrt{\frac{2(z-1)}{z}}u^{-z}\ud t\,,  \label{4.1.11}\\
	W_0&=&\frac{z(d-1)}{L^2}\,,\qquad Z_0=1\,,  \label{4.1.12}\\
	V_0L^2&=&z^2+(d-2)z+(d-1)^2.\label{4.1.13}
\eea
As expected, this coincides with the $(d+1)$-dimensional Lifshitz background of \cite{taylor}. Note that, for $z=1$, the Maxwell field vanishes and one recovers the AdS spacetime.

\subsection{From a higher-dimensional massive $q$-form field\label{app:C.2}}

We start from the higher-dimensional action
\be
\tilde  S=\tilde M^{2}\int d^{p+n+1}x\sqrt{-\tilde g}\left[\tilde R+2\Lambda-\frac1{2(n+2)!}\left(\tilde G_{[n+2]}\right)^2-\frac1{2(n+1)!}\left(\tilde B_{[n+1]}\right)^2\right]
\label{4.2.1}
\ee
where $G_{[n+2]}=\ud B_{[n+1]}$ is a two-form field strength for a Maxwell potential, $\tilde g$ the $(p+n+1)$-dimensional metric and
\be \label{4.2.2}
\left(\tilde G_{[n+2]}\right)^2=\tilde g^{A_1B_1}\ldots\tilde g^{A_{n+2}B_{n+2}}G_{[2]A_1\ldots A_{n+2}}G_{[2]B_1\ldots B_{n+2}}
\ee
\be \label{4.2.3}
\left(\tilde B_{[n+1]}\right)^2=\tilde g^{A_1C_1}\ldots\tilde g^{A_{n+1}C_{n+1}}B_{[1]A_1\ldots A_{n+1}}B_{[1]C_1\ldots C_{n+2}}
\ee
Then, using the reduction Ansatz
\be
	\ud \tilde s^2 = e^{-\da\phi}\ud s^2+ e^{\frac{\phi}{\da}\left(\frac2{p-1}-\da^2\right)}\ud R^2_{(n)} \,,\label{4.2.4}
\ee
for the metric with
\be
	\da^2={2\over p-1}~\frac{n}{(p+n-1)}\leq \da_c^2={2\over p-1}\,,
\label{4.2.5}\ee
and
\be \label{4.2.6}
	G_{[n+2]}=F_{[2]}\wedge \ud R_{(n)}\,,\qquad B_{[n+1]}=A_{[1]}\wedge \ud R_{(n)}\,,
\ee
for the form fields,
one obtains for the gauge field quadratic invariants
\be
	\left(\tilde G_{[n+2]}\right)^2 = \frac{(n+2)!}{2}e^{-(p-3)\da\phi}\left(F_{[2]}\right)^2,\qquad \left(\tilde B_{[n+1]}\right)^2=(n+1)!e^{-(p-2)\da\phi}\left(A_{[1]}\right)^2. \label{4.2.7}
\ee
Combining with
\be
	\sqrt{-\tilde g}=\sqrt{-g}e^{-\da\phi}, \label{4.2.8}
\ee
the reduced action is
\be
 S=M^{2}\int d^{p+1}x\sqrt{- g}\left[ R-\frac12\partial\phi^2+2\Lambda e^{-\da\phi}-\frac14e^{-(p-2)\da\phi}\left( F_{[2]}\right)^2-\frac12e^{-(p-1)\da\phi}\left( A_{[1]}\right)^2\right]
\label{4.2.9}
\ee
which has
\be
	\ga=-(p-2)\da\,,\qquad \epsilon =-(p-1)\da\,,\qquad \ga-\da=\epsilon \label{4.2.10}
\ee
and so verifies the relations $\ga-\da=\epsilon$ for any dimension $p$ and in particular for $p=3$.

Again, replacing in \eqref{MassiveSolRunningScalar}, rescaling the coordinates and setting $n=d-3$, the higher-dimensional solution reads,
\bea
	\tilde \ud s^2&=&L^2{\ud u^2\over u^2}+{\ud x^2+\ud y^2\over u^2}+{-\ud t^2+\ud R_{(d-3)}^2\over u^{2z}}\,,\label{4.2.13}\\
	B_{[d-2]}&=&\sqrt{\frac{2(z-1)}{(d-2)z}}e^{-\sqrt{\frac{d-3}{d-1}}\phi_0}u^{-(d-2)z}\ud t\wedge\ud R_{(d-3)}\,,\label{4.2.14}\\
	W_0&=&\frac{2(d-2)z}{L^2}\,,\quad Z_0=1\,,\label{4.2.15}\\
	V_0L^2&=&(d-2)^2z^2+(3d-8)z+4\,,\label{4.2.16}
\eea
which is a solution of an Einstein AdS theory with a massive $(d-2)$-form, \eqref{4.2.1}. For the special case $d=4$, that is with a massive $2$-form, this coincides with the result of \cite{taylor}. Note that, like in the paradigmatic case of asymptotically flat $\Lambda=0$ black $(d-3)$-branes, the form is supported by a $(d-3)$-dimensional torus. For $z=1$, the form vanishes and one recovers the AdS spacetime.

\subsection{Including a higher-dimensional scalar}

The two previous reductions produced a one-parameter family of uplifts, since only the exponent $\da$ could be chosen independently. One can do a little better and obtain a two-parameter of uplifts, by including a real scalar in the higher-dimensional setup, both in the case of a Maxwell field and an $(n+2)$-form.

\subsubsection{And a Maxwell field}

 We now follow section 7 of \cite{gk}, and include a scalar field $\Phi$ in the higher-dimensional theory. This allows to avoid fixing $\g$ proportional to $\d$. Start from
\be \label{1031}
	 S=\int\ud^{p+n+1}x\sqrt{-g}\left[R-\half\partial\Phi^2-\frac{Z_0}4e^{\Gamma\Phi}F^2-\frac{W_0}2e^{E\Phi}A^2+V_0\right],
\ee
and reduce along
\be \label{1032}
	\ud s^2 = e^{-\Delta\varphi}\ud s^2_{(p+1)}+e^{\frac{2\varphi}{(p-1)\Delta}\left(1-\frac{p-1}2\Delta^2\right)}\ud R^2_{(n)},\quad \frac{p-1}2\Delta^2=\frac{n}{p+n-1}\,.
\ee
Then, the reduced action is
\be \label{1033} S=\int\ud^{p+1}x\sqrt{-g}\left[R-\half\partial\Phi^2-\half\partial\varphi^2-\frac{Z_0}4e^{\Gamma\Phi+\Delta\varphi}F^2-\frac{W_0}2e^{E\Phi}A^2+V_0 e^{-\Delta\varphi}\right],
\ee
from which we can derive the field equations for the two scalars
\bea
	\square \Phi&=&\frac{\Gamma Z_0}4e^{\Gamma\Phi+\Delta\varphi}F^2+\frac{EW_0}2e^{E\Phi}A^2 \label{1034}\\
	 \square\varphi&=&\frac{\Delta Z_0}4e^{\Gamma\Phi+\Delta\varphi}F^2+\Delta V_0 e^{-\Delta\varphi}.\label{1035}
\eea
We would like to truncate to a single scalar theory, and to that end, we set
\be \label{1036}
	\Phi=\alpha\varphi\,,
\ee
which, combining \eqref{1034} and \eqref{1035}, implies that
\be \label{1037}
	0= \frac{Z_0}4\left(\Gamma-\alpha\Delta\right)^2 F^2e^{(\alpha\Gamma+\Delta)\varphi}+\frac{EW_0}2e^{\alpha E\varphi}A^2 - \alpha\Delta V_0 e^{-\Delta\varphi}.
\ee
Setting
\be \label{1038}
	\varphi=\frac{\phi}{\sqrt{1+\a^2}}\,,\quad \delta=\frac{\Delta}{\sqrt{1+\a^2}}\,,\quad \g=\frac{\a\Gamma+\Delta}{\sqrt{1+\a^2}}\,,\quad \e=\frac{\alpha E}{\sqrt{1+\a^2}}
\ee
allows to transform the consistency equation \eqref{1037} to
\be \label{1039}
	0= \frac{Z_0}4\left(\g-\d-\alpha^2\d\right) F^2e^{\g\phi}+\frac{\e W_0}2e^{\e\phi}A^2 - \alpha^2\d V_0 e^{-\d\phi},
\ee
as well as the action \eqref{1033}
\be \label{1040}
S=\int\ud^{p+1}x\sqrt{-g}\left[R-\half\partial\phi^2-\frac{Z_0}4e^{\gamma\phi}F^2-\frac{W_0}2e^{\e\phi}A^2+V_0 e^{-\d\phi}\right],
\ee
which was what we aimed at. Moreover, note that setting $\e=\g-\d$ in the lower-dimensional theory implies that $E=\Gamma$ in the higher-dimensional theory from \eqref{1038}, which is its direct transcription in terms of higher-dimensional variables.

Solving \eqref{1039} for $\a$ after replacing with the solution \eqref{MassiveSolRunningScalar}, we find that
\be
	\alpha^2=\frac{\theta}{\d^2(\theta-2)}-1\,.
\ee
The higher-dimensional solution is Lifshitz with a massive gauge field and a running scalar:
\bea
	\ud s^2&=&-\frac{\ud t^2}{\rho^{2z}}+\frac{\ud R^2_{(n+2)}+\tilde L^2\ud \rho^2}{\rho^2}\,,\\
	e^\Phi&=&\rho^{\Gamma(z-1)}\,,\quad \tilde L^2=\frac{Z_0}{4W_0}\left[2(n+2)+\Gamma^2(z-1)\right]\left[2z+\Gamma^2(z-1)\right],\\
	A_t&=&Q\rho^{-z+\frac{\Gamma}{2}(z-1)}\,,\qquad Q^2=\frac{4(z-1)}{Z_0\left[2z+(z-1)\Gamma^2\right]}
\eea
Moreover, we find that $\theta=-n<0$, and $d_{eff}=2+n$. For $\Gamma=0$, we recover the solution without running scalar of section \ref{app:C.1}.

\subsubsection{And a $(q+1)$-form field}

We start from the higher-dimensional action
\be
\tilde  S=\tilde M^{2}\int d^{p+n+1}x\sqrt{-\tilde g}\left[\tilde R+V_0-\frac12\partial\Phi^2-\frac{Z_0e^{\Gamma\Phi}}{2(n+2)!}\left(\tilde G_{[n+2]}\right)^2-\frac{W_0e^{E\Phi}}{2(n+1)!}\left(\tilde B_{[n+1]}\right)^2\right]
\label{C.40}
\ee
where $G_{[n+2]}=\ud B_{[n+1]}$ is a two-form field strength for a Maxwell potential, $\tilde g$ the $(p+n+1)$-dimensional metric and
\be \label{c.41}
\left(\tilde G_{[n+2]}\right)^2=\tilde g^{A_1B_1}\ldots\tilde g^{A_{n+2}B_{n+2}}G_{[2]A_1\ldots A_{n+2}}G_{[2]B_1\ldots B_{n+2}}
\ee
\be \label{C.42}
\left(\tilde B_{[n+1]}\right)^2=\tilde g^{A_1C_1}\ldots\tilde g^{A_{n+1}C_{n+1}}B_{[1]A_1\ldots A_{n+1}}B_{[1]C_1\ldots C_{n+2}}
\ee
\be \label{C.43}
	\ud s^2 = e^{-\Delta\varphi}\ud s^2_{(p+1)}+e^{\frac{2\varphi}{(p-1)\Delta}\left(1-\frac{p-1}2\Delta^2\right)}\ud R^2_{(n)},\quad \frac{p-1}2\Delta^2=\frac{n}{p+n-1}\,.
\ee
Then, the reduced action is
\be \label{C.44} S=\int\ud^{p+1}x\sqrt{-g}\left[R-\half\partial\Phi^2-\half\partial\varphi^2-\frac{Z_0}4e^{\Gamma\Phi-(p-2)\Delta\varphi}F^2-\frac{W_0}2e^{E\Phi-(p-1)\Delta\varphi}A^2+V_0 e^{-\Delta\varphi}\right],
\ee
from which we can derive the field equations for the two scalars
\be\label{C.45}
\begin{split}
\square \Phi&=\frac{\Gamma Z_0}4e^{\Gamma\Phi-(p-2)\Delta\varphi}F^2+\frac{EW_0}2e^{E\Phi-(p-1)\Delta\varphi}A^2 \\
	 \square\varphi&=\Delta V_0 e^{-\Delta\varphi}-\frac{(p-2)\Delta\varphi Z_0}4e^{\Gamma\Phi-(p-2)\Delta\varphi}F^2-\frac{(p-1)\Delta W_0}{2}e^{E\Phi-(p-1)\Delta\varphi}A^2.
\end{split}
\ee
We would like to truncate to a single scalar theory, and to that end, we set
\be \label{C.47}
	\Phi=\alpha\varphi\,,
\ee
which, combining the two equations in \eqref{C.45}, implies that
\be \label{C.48}
	0= \frac{Z_0}4\left(\Gamma+(p-2)\alpha\Delta\right) F^2e^{(\alpha\Gamma-(p-2)\Delta)\varphi}+\frac{\left(E+(p-1)\alpha\Delta\right)W_0}2e^{(\alpha E-(p-1)\Delta)\varphi}A^2 - \alpha\Delta V_0 e^{-\Delta\varphi}.
\ee
Setting
\be \label{C.49}
	\varphi=\frac{\phi}{\sqrt{1+\a^2}}\,,\quad \delta=\frac{\Delta}{\sqrt{1+\a^2}}\,,\quad \g=\frac{\a\Gamma-(p-2)\Delta}{\sqrt{1+\a^2}}\,,\quad \e=\frac{\alpha E-(p-1)\Delta}{\sqrt{1+\a^2}}
\ee
allows to transform the consistency equation \eqref{C.48} to
\be \label{C.50}
	0= \frac{Z_0}4\left(\g+(p-2)(1+\alpha^2)\d\right) F^2e^{\g\phi}+\frac{\left(\e+(p-1)(1+\al^2)\d \right)W_0}2e^{\e\phi}A^2 - \alpha^2\d V_0 e^{-\d\phi},
\ee
as well as the action \eqref{C.44}
\be \label{C.51}
S=\int\ud^{p+1}x\sqrt{-g}\left[R-\half\partial\phi^2-\frac{Z_0}4e^{\gamma\phi}F^2-\frac{W_0}2e^{\e\phi}A^2+V_0 e^{-\d\phi}\right],
\ee
which was what we aimed at. Moreover, note that setting $\e=\g-\d$ in the lower-dimensional theory implies that $E=\Gamma$ in the higher-dimensional theory from \eqref{C.49}, which is its direct transcription in terms of higher-dimensional variables.

Solving \eqref{C.50} for $\a$ after replacing with the solution \eqref{MassiveSolRunningScalar}, we find that
\be
	\alpha^2=\frac{\theta}{\d^2(\theta-2z)}-1\,.
\ee
The higher-dimensional solution is a Lifshitz $n$-brane with a massive $(n+1)$-potential and a running scalar:
\bea
	\ud s^2&=&-\frac{\ud t^2+\ud R^2_{(n)}}{\rho^{2z}}+\frac{\ud R^2_{(2)}+ L^2\ud \rho^2}{\rho^2}\,,\quad e^\Phi=\rho^{-\Gamma(z-1)}\,,\\
	L^2&=&\frac1{2V_0}\left[8+2(3n+1)z+2(n+1)^2z^2+\Gamma^2(z-1)^2\right],\\
	B_{[n+1]}&=&Q\rho^{-(n+1)z-\frac{\Gamma^2}{2}(z-1)}\ud t\wedge\ud R_{[n]}\,,\quad Q^2=\frac{4(z-1)}{Z_0\left[2(n+1)z+(z-1)\Gamma^2\right]}\\
	 \frac{W_0}{Z_0V_0}&=&\frac{\left[4+(1-z)\Gamma^2\right]\left[2(n+1)z-(1-z)\Gamma^2\right]}{2\left(8+2(3n+1)z+2(n+1)^2z^2+\Gamma^2(z-1)^2\right)}
\eea
Moreover, we find that $\theta=-nz<0$, and $d_{eff}=2+nz$. For $\Gamma=0$, we recover the solution without running scalar of section \ref{app:C.2}.

\end{document}